\documentclass[sigconf]{acmart}

\usepackage{booktabs} % For formal tables
\usepackage{float}
\usepackage{arydshln} % for dashed lines in tables (tabu is much more flexible)
\usepackage{enumitem}
\usepackage{natbib}
\usepackage{graphicx}
\graphicspath{{data/}}
\usepackage{pdfpages}
\usepackage{nicefrac}
\usepackage{blindtext}
\usepackage{tabularx}
\usepackage{caption}
\usepackage{subcaption}
\usepackage{bm} % produces better bold symbols than does \boldsymbol{x}; use \bm{x}
\usepackage{amsmath, mathrsfs,amsfonts,amssymb}
\usepackage{wasysym} % smiley face
\usepackage{xfrac} % use command \sfrac{1}{2} to get a nice 1/2 symbol.
\usepackage{savesym} % openbox symbol conflict between tx fonts and amsthm
\savesymbol{openbox}
\usepackage{amsthm} %theorem like environments
\usepackage{thmtools}
\usepackage{cleveref}
\usepackage{commath}
\usepackage[export]{adjustbox}
\usepackage{dblfloatfix}
\usepackage{tikz}
\usepackage{blindtext}
\usepackage{xcolor}
\usepackage{url}
\usetikzlibrary{backgrounds}

\definecolor{lightgray}{HTML}{989898}
\definecolor{darkgray}{HTML}{696969}

\newcommand{\lightgraybg}[1]{%
  \begingroup\setlength{\fboxsep}{3pt}% no padding
  \raisebox{2pt}{\colorbox{lightgray}{#1}}
  \endgroup
}

\newcommand{\darkgraybg}[1]{%
  \begingroup\setlength{\fboxsep}{3pt}% no padding
  \raisebox{2pt}{\colorbox{darkgray}{#1}}
  \endgroup
}

\newcommand{\psame}{p_{\scriptsize\mbox{same}}}
\newcommand{\pdiff}{p_{\scriptsize\mbox{diff}}}
\newcommand{\pjump}{p_{\scriptsize\mbox{jump}}}
\newcommand{\pout}{p_{\scriptsize\mbox{out}}}
\newcommand{\plink}{p_{\scriptsize\mbox{link}}}
\newcommand{\rlocal}{r_{\scriptsize\mbox{local}}}

\crefname{subsection}{subsection}{subsections}

\setcopyright{none}

\acmYear{2018}
\copyrightyear{2016}
\renewcommand\footnotetextcopyrightpermission[1]{} % removes footnote with conference information in first column
\begin{document}
\title{Growing Attributed Networks through Local Processes}

\author{Harshay Shah, Suhansanu Kumar, Hari Sundaram}
\affiliation{
  \institution{
  Department of Computer Science \\
  University of Illinois at Urbana-Champaign}
}
\email{{hrshah4, skumar56, hs1}@illinois.edu}

\renewcommand{\shortauthors}{Shah et al.}

\begin{abstract}
%!TEX root = ../draft.tex

This paper proposes an attributed network growth model. Despite the knowledge
that individuals use limited resources to form connections to similar others, we
lack an understanding of how local and resource-constrained mechanisms explain
the emergence of rich structural properties found in real-world networks. We
make three contributions. First, we propose an interpretable and accurate model
of attributed network growth that jointly explains the emergence of in-degree
distribution, local clustering, clustering-degree relationship and attribute
mixing patterns. Second, we make use of biased random walks to develop a model
that forms edges locally, without recourse to global information. Third, we
account for multiple sociological phenomena---bounded rationality; structural
constraints; triadic closure; attribute homophily; preferential attachment.
We explore the parameter space of the proposed Attributed Network Growth (\texttt{ARW})
to show each model parameter intuitively modulates network structure.
Our experiments show that \texttt{ARW} accurately preserves network structure and attribute mixing patterns of
six real-world networks; it improves upon the performance of eight
well-known models by a significant margin of
2.5--$10\times$.

% However, well-known growth models that preserve multiple structural properties do not
% incorporate these resource constraints. Conversely, resource constrained growth models
% cannot jointly preserve multiple structural
% properties of real networks. Furthermore, most growth models disregard
% the effect of homophily on edge formation and global network structure.
% Our Attributed Random Walk (\texttt{ARW}) model explains how structural \&
% content-based properties of real-world networks jointly arise from individual
% preferences \& edge formation under constraints of limited information and network access.

% In our model, each node that joins the network selects a seed node from which it initiates a
% biased random walk to concurrently explore the network and link to existing nodes.
% At each step of the walk, the new node either jumps back to
% the seed node or chooses an outgoing or incoming edge to visit another node; It
% probabilistically links to each visited node and halts after forming a few edges
% Through our experiments, we observe that the proposed model \texttt{ARW} preserves
% network structure and attribute mixing patterns of six real-world networks
\end{abstract}

% \keywords{Network growth models, Attributed networks, Homophily}

\maketitle

%!TEX root = ../draft.tex

% datasets table..
\begin{table*}[b]
 {
  \begin{tabular}[c]{llrrrc|rrrr} \toprule
   Network &  Description & $|V|$           & $|E|$        & $T$        & $A$, $|A|$  &  \texttt{LN} $(\mu, \sigma)$ & \texttt{DPL} $\alpha$       &  Avg. ${\texttt{LCC}}$  & \texttt{AA} $r$           \\ \midrule
   \texttt{USSC}~\cite{fowler2008authority}  & U.S. Supreme Court cases         & 30,288     & 216,738      & 1754-2002  & -   & (1.19, 1.18) & 2.32     & 0.12    & -     \\
   \texttt{HEP-PH}~\cite{gehrke2003overview} & ArXiv Physics manuscripts     & 34,546     & 421,533      & 1992-2002  & -  &   (1.32, 1.41) & 1.67     & 0.12    & -                 \\
   \texttt{Semantic}~\cite{ammar}&   Academic Search Engine  & 7,706,506  & 59,079,055   & 1991-2016  & -   &   (1.78, 0.96)  & 1.58     & 0.06    & -             \\   \midrule
   \texttt{ACL}~\cite{acldata}    & NLP papers      & 18,665     & 115,311      & 1965-2016  & \textsc{venue}, 50  &   (1.93, 1.38)  & 1.43     & 0.07    & 0.07  \\
   \texttt{APS}~\cite{aps}     & Physics journals     & 577,046    & 6,967,873    & 1893-2015  & \textsc{journal}, 13 &   (1.62, 1.20)  & 1.26     & 0.11    & 0.44 \\
   \texttt{Patents}~\cite{leskovec2005graphs}   & U.S. NBER patents    & 3,923,922  & 16,522,438   & 1975-1999  & \textsc{category}, 6 &   (1.10, 1.01)   & 1.94     & 0.04    & 0.72 \\
   % \texttt{PYPI}         & 25,169     & 71,371       & 2002-2018  & \textsc{category} & 9  \\
  \bottomrule
  \end{tabular}
  \vspace{1mm}
  \caption{Summary statistics \& global properties of six network datasets: $|V|$ nodes join the networks and form edges $|E|$ over
  time period $T$. In attributed networks, each node has a categorical attribute value that belongs to set $A$ of size $|A|$.
  The networks exhibit lognormal (\texttt{LN}) in-degree distribution with mean $\mu$ and standard deviation $\sigma$,
  high average local clustering (${\texttt{LCC}}$) \& attribute assortativity (\texttt{AA}) coefficient and
  densify over time with power law (\texttt{DPL}) exponent $\alpha$.}
  \label{table:datasets}
  \label{table:netstats}
 }
\end{table*}

\section{Introduction}
\label{sec:Introduction}

% what is the problem?

We present a network growth model that explains how distinct
structural properties of attributed networks can emerge from local edge
formation processes. In real-world networks, individuals tend to form edges
despite limited information and partial network access.
Moreover, phenomena such as triadic closure and homophily
\textit{simultaneously} influence individuals' decisions to form connections.
Over time, these decisions cumulatively shape real-world networks to exhibit
rich structural properties: heavy-tailed in-degree distribution, skewed
local clustering and homophilic mixing patterns. However, we lack an
understanding of local, resource-constrained mechanisms that incorporate
sociological factors to explain the emergence of rich structural properties.

% The problem of developing a model of network growth, where individuals act under
% resource constraints, including access to only local information is hard. The
% problem lies in identifying simple mechanisms with few parameters that unifies
% multiple sociological phenomena to \textit{jointly} preserve structural
% properties and attribute mixing patterns of attributed networks.

% why is it important?

Classic models of network growth tend to make unrealistic assumptions about how
individuals form edges. Consider a simple stylized example: the process of
finding a set of papers to cite when writing an article. In preferential
attachment \cite{barabasi1999emergence} or fitness
\cite{bianconi2001bose,caldarelli2002scale,wang2013quantifying} based models, a
node making $m$ citations would pick papers from the \textit{entire} network in
proportion to their in-degree or fitness respectively. This process assumes that
individuals possess {complete} knowledge of in-degree or fitness of every node
in the network. An equivalent formulation---vertex copying
\cite{kumar2000stochastic}---induces preferential attachment: for every
citation, a node would pick a paper uniformly at random from \textit{all}
papers, and either cite it or copy its citations. Notice that the copying
mechanism assumes individuals have complete access to the network and forms each
edge independently. Although these models explain the emergence of power law
degree distributions, they are unrealistic: they require global knowledge (e.g.,
preferential attachment requires knowledge of the global in-degree distribution)
or global access (e.g., vertex copying requires random access to all nodes).
Additionally, these models do not account for the fact that many
networks are attributed (e.g., a paper is published at a venue; a Facebook user
may use gender, political interests to describe them) and that assortative
mixing is an important network characteristic~\cite{newman2002assortative}.

Recent papers tackle resource constraints
\cite{mossa2002truncation,zeng2005construction,wang2009local} as well as nodal
attributes \cite{de2013scale,gong2012evolution}. However, the former disregard
attributes and the latter do not provide a realistic representation of edge
formation under resource constraints. Furthermore, both sets of models do not
jointly preserve multiple structural properties. Developing an interpretable and
accurate model of attributed network growth that accounts for observed
sociological phenomena is non-trivial. Accurate network growth models are useful
for synthesizing networks as well as to extrapolate existing real-world
networks.

% representations of how individuals make decisions about edge formation. A
% realistic representation of how individuals form edges necessitates modeling the
% effect of multiple sociological phenomena on edge formation under resource
% constraints.

%why is it hard?

% what did we do?
\begin{figure}[t]
	\centering
	\includegraphics[width=\columnwidth]{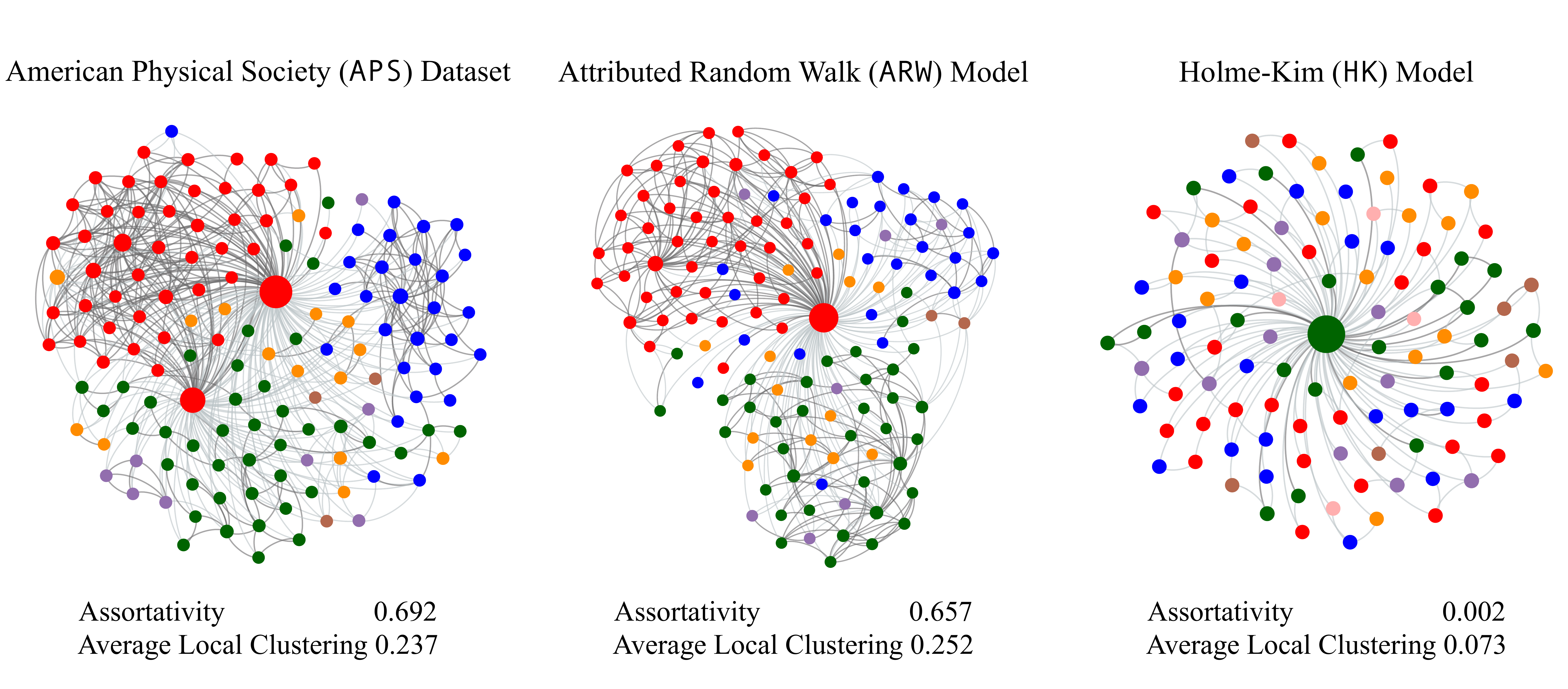}
    \vspace{1pt}
    \caption{The figure shows how our proposed model of an Attributed Random Walk (\texttt{ARW}) accurately preserves local clustering and assortativity; we contrast with a non-attributed growth model~\cite{holme2002growing} to underscore the importance of using attributes for network growth.}
	\label{fig:intro_plot}
\end{figure}

We aim to develop a growth model that accounts for resource constraints and sociological phenomena
influencing edge formation in addition to preserving global network structure.
We make three key contributions.
First, we propose a simple and accurate model of attributed network growth.
Second, our model is based on local processes to form edges, without recourse to global network information.
Third, our model unifies multiple sociological phenomena---bounded rationality; structural constraints; triadic closure; attribute homophily; preferential attachment---to jointly  model global network structure and attribute mixing patterns.

The proposed model---Attributed Random Walk (\texttt{ARW})---jointly explains
the emergence of in-degree distribution, local clustering, clustering-degree
relationship and attribute mixing patterns through a resource constrained
mechanism based on random walks (see~\Cref{fig:intro_plot}). In particular,
the model relies entirely on local information to grow the network, without
access to information of all nodes. In \texttt{ARW}, incoming nodes select a
seed node based on attribute similarity and initiate a biased random walk: at
each step of the walk, the incoming node either jumps back to its seed or
chooses an outgoing link or incoming link to visit another node; it links to
each visited node with some probability and halts after it has exhausted its
budget to form connections.
Our experiments on six large-scale network datasets indicate that the proposed growth model outperforms
eight state-of-the-art network growth models by a
statistically significant margin of 2.5--$10\times$.
Furthermore, we analyze the parameter space of the model to show how each parameter intuitively controls
one or more key structural properties.

The rest of the paper is organized as follows.
We begin by defining the problem statement in~\Cref{sec:Problem Statement}.
In~\Cref{sec:Analysis}, we outline six network datasets, describe key structural
properties of real-world networks and discuss insights from sociological studies.
Then, in~\Cref{sec:Proposed Model}, we describe the network growth model. We follow
by presenting experiments in ~\Cref{sec:Experiments}, analysis of assortative mixing
in ~\Cref{subsec:LocalMixing} and discussion in~\Cref{sec:Discussion}.

%!TEX root = ../draft.tex

\section{Problem Statement}
\label{sec:Problem Statement}

Consider an attributed directed network $G=(V,E,B)$, where $V$ \& $E$ are
sets of nodes \& edges and each node has an attribute value $b \in B$.
The goal is to develop a directed network growth model that preserves structural
and attribute based properties observed in $G$. The growth model should be
normative, accurate and parsimonious:
\begin{enumerate}
\item \textbf{Normative}: The model should account for multiple sociological phenomena that influence how individuals form edges under constraints of limited global information and partial network access.

\item \textbf{Accurate}: The model should preserve key structural
and attribute based properties: degree distribution,
local clustering, degree-clustering relationship and attribute mixing patterns.

\item \textbf{Parsimonious}: The model should be able to
generate networks with tunable structural properties, while having few parameters.
\end{enumerate}

Next, we present empirical analysis on real-world datasets to motivate our attributed random walk model.

% \clearpage
%!TEX root = ../draft.tex

\section{Empirical Analysis}
\label{sec:Analysis}

We begin by describing six large-scale network datasets that we
use in our analysis and experiments. Then, we describe global network properties,
insights from empirical studies in the Social Science and common assumptions in
network modeling. Finally, we discuss the role of structural proximity in edge
formation.

\subsection{Datasets}
\label{sec:Datasets}

We consider six citation networks of different scales (size, time) from diverse
sources: research articles, utility patents and judicial cases. ~\Cref{table:datasets} lists their
summary statistics and global network properties.  Three of the six datasets are attributed networks;
that is, each node has a categorical attribute value.

We focus on citation networks for two reasons. First, since nodes in citation networks form
all outgoing edges to existing nodes at the time of joining the network,
these datasets provide a clean basis to study edge formation in
attributed networks. Second, the node-level, temporal information in datasets that span long time periods (e.g. \texttt{USSC})
enables us to study structural properties at different time stages via network snapshots.
Next, we study the structural and attribute properties of these networks.

\subsection{Global Network Properties}
\label{subsec:factors}

% Factors that influence edge formation at the nodal level have a cumulative
% effect on global structural properties of real-world networks.
Compact statistical descriptors of global network properties ~\cite{newman2010networks}
such as degree distribution, local clustering, and attribute assortativity
quantify the extent to which local edge formation phenomena shape global network
structure.

\textbf{Degree distribution:}
Real-world networks tend to exhibit heavy tailed degree distributions in which
a small but significant fraction of the nodes turn into high-degree hubs.
We observe that Log-normal fits, with parameters listed in~\Cref{table:netstats}, well describe
the in-degree distributions, consistent with Broido and
Clauset's~\cite{broido2018scale} observation that scale-free, real-world networks
are rare.

% Real-world networks tend to exhibit heavy tailed degree distributions.
% These distributions can emerge from the well-known preferential attachment
% process~\cite{simon1955class,barabasi1999emergence}, where incoming nodes
% connect with nodes in proportion to their degree.
% Over time, preferential
% attachment amplifies initial differences in node degree, giving rise to heavy tailed
% distributions.
% (also known as the ``rich get richer'' effect) in citation networks; It also implies that most papers receive zero or a few
% citations, but a small but significant percent of the nodes turn into popular
% hubs that receive many citations.
% Log-normal fits, with parameters listed in~\Cref{table:netstats}, well describe
% the in-degree distribution of all network datasets, consistent Broido and
% Clauset's~\cite{broido2018scale} observation that scale-free, real-world networks
% are rare
% with truly
% power law degree distributions are rare.
% Our model explains the emergence of heavy tailed in-degree distributions through a
% \textit{local} process that adjusts bias towards linking to well-connected nodes

\begin{figure}[b]
 \vspace{-10pt}
 \centering
 \includegraphics[width=\columnwidth]{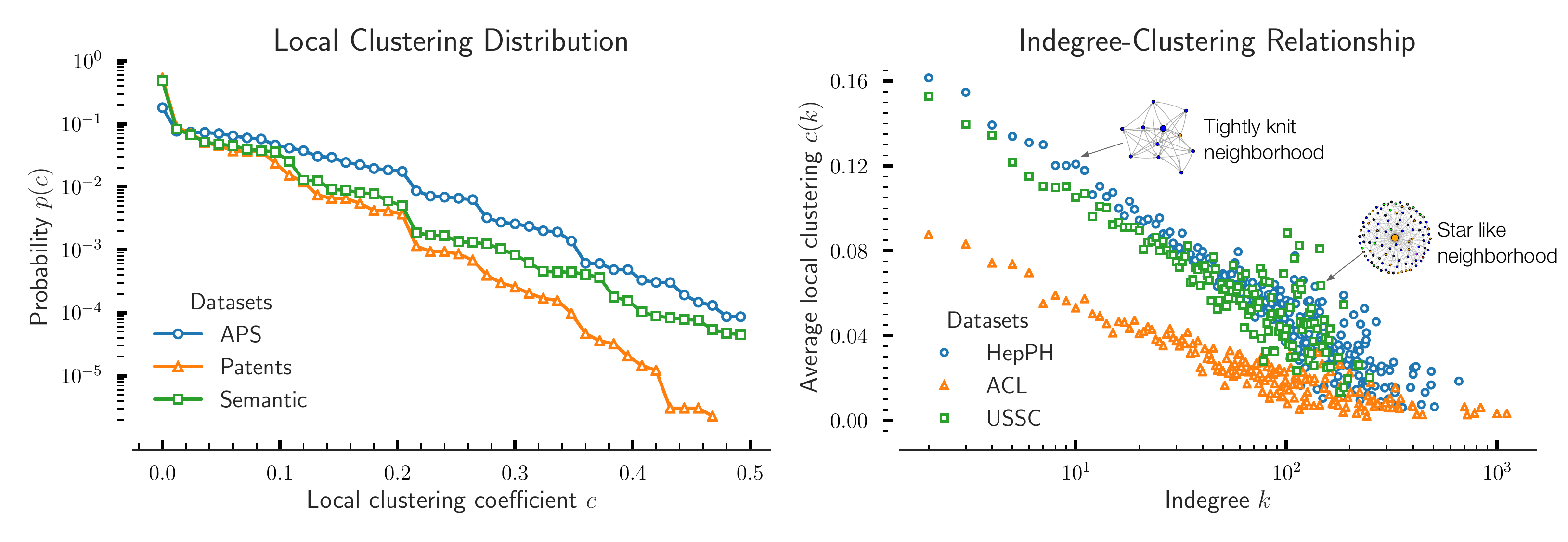}
 \caption{
    Local clustering in real-world networks have common characteristics:
    skewed local clustering distribution (left subplot) and a negatively correlated
    relationship between in-degree and average local clustering (right subplot).
 }
 \label{fig:cc_dc}
 \vspace{-10pt}
\end{figure}

\textbf{Local Clustering:}
Real-world networks exhibit high local clustering
(\texttt{LCC}), as shown in~\Cref{table:netstats}. Local
clustering can arise from triadic closure~\cite{simmel1950sociology,
newman2001clustering}, where nodes with common neighbor(s) have an increased
likelihood of forming a connection.
The coefficient of node $i$ equals the probability with which two randomly chosen
neighbors of the node $i$ are connected. In directed networks, the neighborhood
of a node $i$ can refer to the nodes that link to $i$, nodes that
$i$ links to or both. We define the neighborhood to be the set
of all nodes that link to node $i$. In ~\Cref{fig:cc_dc}, we show that (a) average local clustering is not a
representative statistic of the skewed local clustering distributions and (b) real-world networks
exhibit a negative correlation between in-degree and clustering.
That is, low in-degree nodes have small, tightly knit neighborhoods
and high in-degree nodes tend have large, star-shaped neighborhoods.

\begin{figure}
 \centering
 \includegraphics[width=\columnwidth]{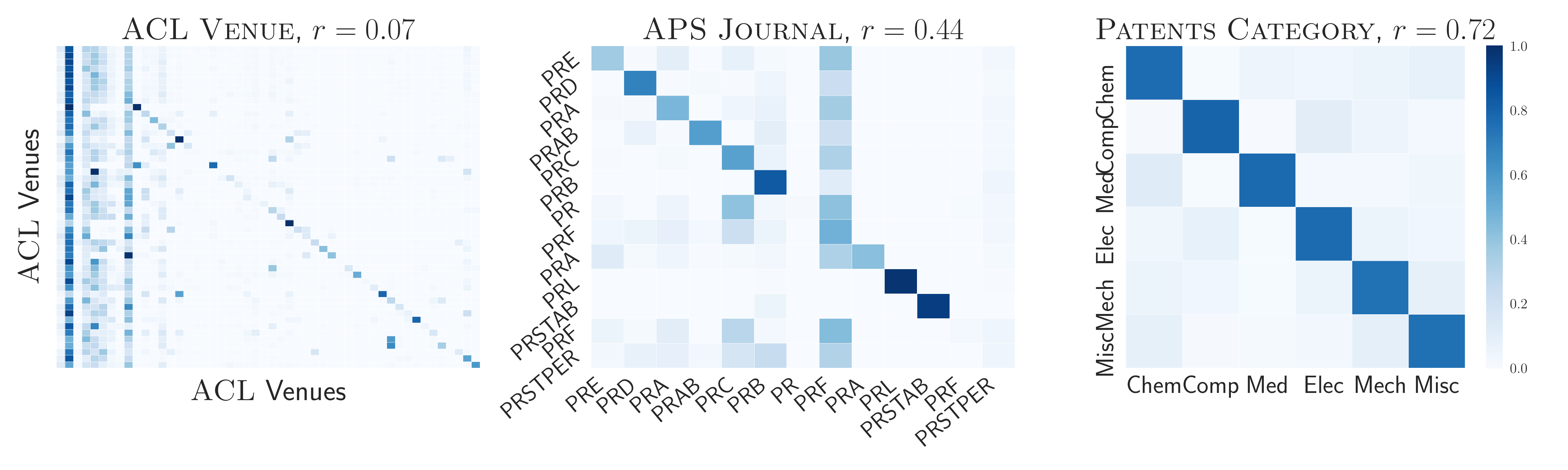}
 \caption{
    Attributed networks \texttt{ACL}, \texttt{APS} and \texttt{Patents} exhibit
    homophily w.r.t attributes \texttt{Venue} ($r=0.07$), \texttt{Journal} ($r=0.44$) and
    \texttt{Category} ($r=0.72$) respectively.
 }
 \label{fig:mixing}
 \vspace{-20pt}
\end{figure}

\textbf{Homophily:}
Attributed networks tend to exhibit homophily~\cite{mcpherson2001birds}, the
phenomenon where similar nodes are more likely to be connected than dissimilar
nodes. The assortativity coefficient ~\cite{newman2002assortative} $r \in [-1,
1]$, quantifies the level of homophily in an attributed network.
Intuitively,
assortativity compares the observed fraction of edges between nodes with the same attribute
value to the expected fraction of edges between nodes with same attribute value
if the edges were rewired randomly. In~\Cref{fig:mixing}, we show that
attributed networks \texttt{ACL}, \texttt{APS} and \texttt{Patents} exhibit
varying level of homophily with assortativity coefficient ranging from $0.07$ to
$0.72$.
% Attributed networks tend to exhibit homophily~\cite{mcpherson2001birds}, the
% phenomenon where similar nodes are more likely to be connected than dissimilar
% nodes. The assortativity coefficient ~\cite{newman2002assortative} $r \in [-1,
% 1]$, quantifies the level of homophily in an attributed network.
% %  and indicates
% % the extent to which attribute similarity influences edge formation.
% Intuitively,
% assortativity compares the observed fraction of edges between nodes with the same attribute
% value to the expected fraction of edges between nodes with same attribute value
% if the edges were rewired randomly. In~\Cref{fig:mixing}, we show that
% attributed networks \texttt{ACL}, \texttt{APS} and \texttt{Patents} exhibit
% varying level of homophily with assortativity coefficient ranging from $0.07$ to
% $0.72$.

% We embed attribute based preferences at the local level lead to generate networks
% with varying attribute mixing patterns.

\textbf{Increasing Out-degree over Time:}
The out-degree of nodes that join real-world networks tends to increase as
functions of network size and time. This phenomenon densifies networks and
can shrink effective diameter over time. Densification tends to exhibit a power law
relationship ~\cite{leskovec2005graphs} between the number of edges $e(t)$ and
nodes $n(t)$ at time $t$: $e(t) \propto n(t)^{\alpha}$.~\Cref{table:netstats}
lists the densification power law (\texttt{DPL}) exponent $\alpha$ of the
network datasets.

To summarize, citation networks tend to be homophilic networks that undergo
accelerated network growth and exhibit regularities in structural properties:
heavy tailed in-degree distribution, skewed local clustering distribution,
negatively correlated degree-clustering relationship, and varying attribute
mixing patterns.

% In our proposed model, we increase the outdegree of incoming nodes at a linear
% or superlinear rate to account for the accelerated network growth observed in
% real networks.

% To summarize, factors such as preferential attachment, triadic closure and
% homophily not only influence how individuals form connections at the local level
% but also explain regularities arise in global structural properties of
% real-world networks. Next, we discuss empirical studies from sociology that
% examine network formation and decision making.

\subsection{Insights from Sociological Studies}
Sociological studies on network formation seek to explain
how individuals form edges in real-world networks.

\textbf{Interplay of Triadic Closure and Homophily:}
Empirical studies~\cite{35626,block2014multidimensional} that analyze the
interplay between triadic closure and homophily
 % in evolving networks
  indicate
that \textit{both} structural proximity and homophily are statistically
significant factors that simultaneously influence edge formation.
Homophilic preferences~\cite{mcpherson2001birds} induce edges between similar
nodes, whereas structural factors such as network distance limit
edge formation to proximate nodes (e.g. friend of a
friend).

\textbf{Bounded Rationality:}
Extensive work~\cite{simon1972theories,gigerenzer1996reasoning,lipman1995information} on
decision making shows that individuals are boundedly rational
actors; constraints such as limited information, cognitive capacity and time impact decision making.
This suggests that resource-constrained individuals that join networks are likely to employ simple rules
to form edges using limited information and partial network access.

Current preferential attachment and fitness-based models
\cite{dorogovtsev2000structure,singh2017relay,barabasi1999emergence}
make two assumptions that are at variance with these findings.
First, by assuming that successive edge formations are independent,
these models disregard the effect of triadic closure and structural proximity.
Second, as discussed in~\cref{sec:Introduction}, these models require complete network
access or knowledge of node-level properties.

Insights from sociological studies indicate that edge formation in real-world
networks comprises biases towards nodes that are similar, well-connected or
structurally proximate. Coupled with empirical analyses, it also motivates
the need to model how resource-constrained edge formation processes collectively
shape global network properties of large-scale networks over time.

\subsection{Proximity-biased Edge Formation}

\begin{figure}[H]
%  \vspace{-4pt}
 \centering
 \includegraphics[width=\columnwidth]{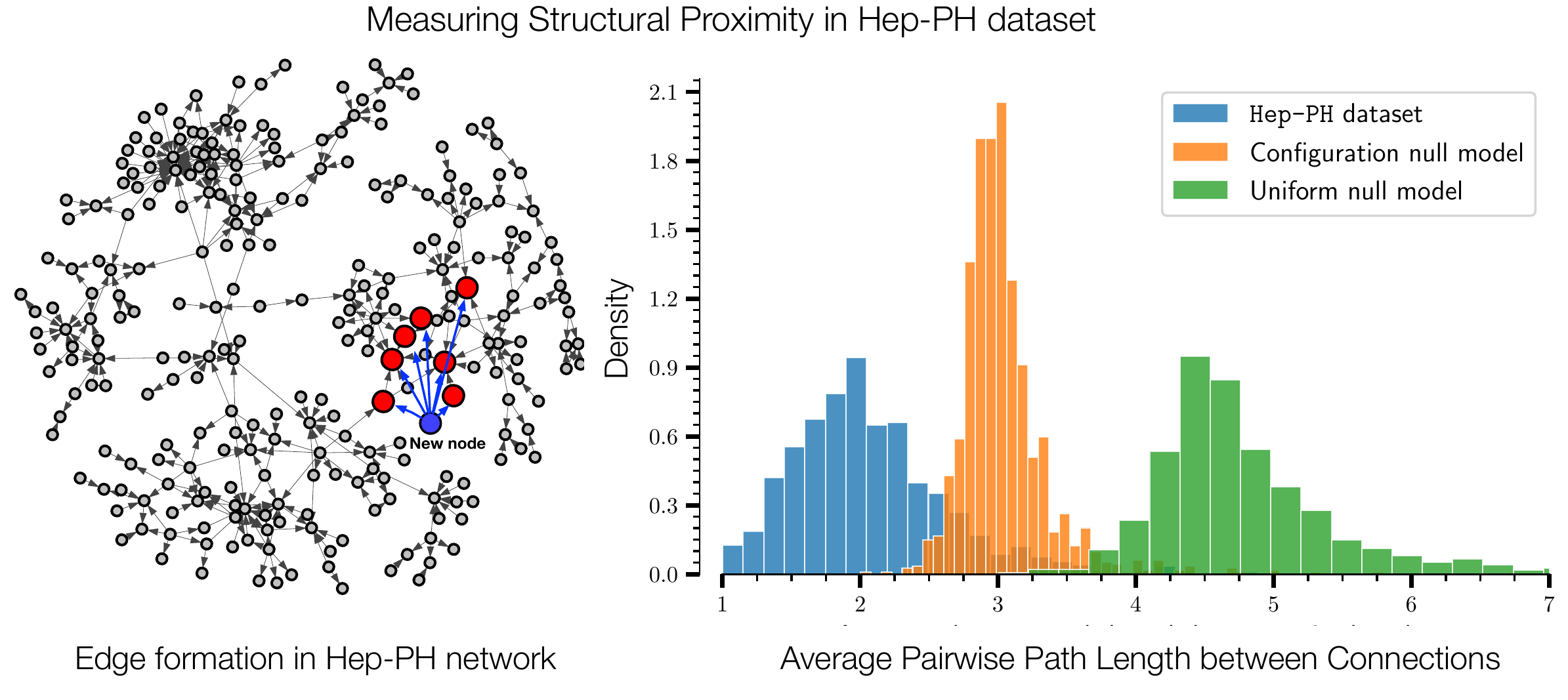}
 \caption{
    Proximity-biased edge formation. The diagram and proximity distributions
    collectively indicate how edge formation in real-world networks are biased
    towards structurally proximate nodes in the same locality.  }
 \label{fig:locality}
 % \vspace{-10pt}
\end{figure}

We investigate the effect of structural proximity on edge formation in real-world networks.
Prior work~\cite{35626} shows that the probability of edge formation in social networks
decreases as a function of network distance.
Indeed, triadic closure explains how individuals form additional
edges to proximate nodes (e.g. friend of friend) over time.
However, we lack a concrete understanding of the extent to which structural
proximity influences edge formation in bibliographic networks, wherein incoming nodes form all edges at
the time of joining the network. In~\Cref{fig:locality}, we show how high structural proximity
among incoming (shown in blue) node's (shown in red) connections in the \texttt{Hep-PH} dataset
hints at edge formation processes biased towards proximate nodes in the same local neighborhood.

% While prior work~\cite{35626} shows that the probability of
% edge formation in social networks decreases as a function of network distance,
% explained through processes like triadic closure, there are is a key difference
% with directed citation networks that we study. In social networks, edge
% formation can take place after the nodes join, whereas in citation networks, a
% node forms all edges at the time of joining. This means that probability of
% triadic closure changes over time. If nodes do not drop out, this probability
% increases with time. Thus in a citation network, it is unclear if one should
% expect if the set of nodes $\{k \}$ that a incoming node $i$ cites should be

% The left subplot in \Cref{fig:locality} shows high structural proximity between
% an incoming (shown in blue) node's  connections (shown in red) in the
% \texttt{Hep-PH} dataset. This hints at edge formation processes biased towards
% nodes in the same local neighborhood.

% Due to the unavailability of navigational data such as the sequence in which individuals form edges,

We rely on network snapshots and
node arrival sequence to estimate a statistic based on path length that measures
structural proximity between nodes' connections. Consider an incoming node $u$ that forms edges to nodes in $N(u)$.
To measure the proximity between node $u$'s connections, we compute the average
pairwise shortest path distance between the connections in the network snapshot
immediately preceding node $u$'s arrival.

The right subplot in~\Cref{fig:locality} compares the proximity statistic
distribution of the \texttt{Hep-PH} dataset to two null models: uniform and
configuration. In the uniform model, incoming nodes form connections to existing
nodes uniformly at random, whereas the configuration model  randomly
rewire all edges in \texttt{Hep-PH} while preserving the out-degree and in-degree
distributions. We first observe that the connections of incoming nodes in the uniform
null model are structurally distant from each other on average.
Although the presence of hubs in the configuration model
considerably decreases the distance between nodes' connections, it does
not explain why the majority of connections in \texttt{Hep-PH} are either
connected directly or via an intermediate node. The disparity between the
observed and null distributions suggests that structural proximity between
connections is intrinsic to edge formation in real-world networks.

To summarize, empirical analyses and insights from the Social Sciences motivate
the need to model how resource-constrained edge formation processes collectively
shape well-defined global network properties of large-scale networks over time.

\begin{figure*}
	\vspace{-15pt}
    \centering
    \includegraphics[width=.9\linewidth]{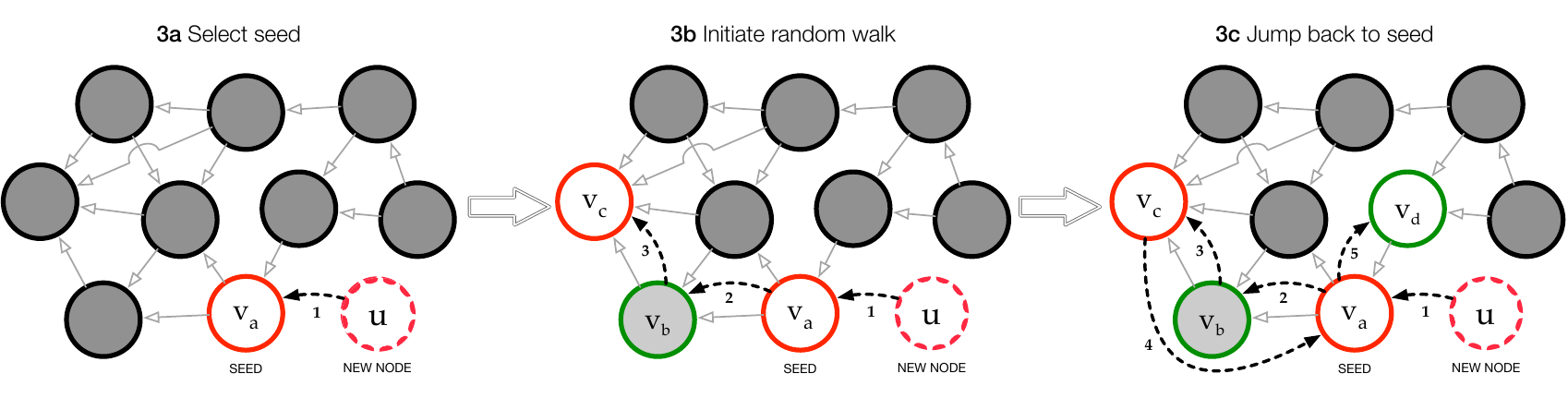}
    \caption{Edge formation in \texttt{ARW}: consider
    an incoming node $u$ with outdegree ${m=3}$ and attribute value {$B(u)=\textsc{red} \in \{\textsc{red},\textsc{green}\}$}.
    In fig. 5a, $u$ joins the network and selects seed $v_a$ via \textsc{Select-Seed}.
    Then, in fig. 5b, $u$ initiates a \textsc{Random-Walk} and traverses from $v_a$ to $v_b$ to $v_c$.
    Finally, $u$ jumps back to its seed $v_a$ and restarts the walk, as shown in fig. 5c.
    Node $u$ halts the random walk after linking to $v_a$, $v_c$ \& $v_d$.
    }
    \label{fig:randomwalk}
	\vspace{-8pt}
\end{figure*}

% \begin{figure*}
% 	\vspace{-20pt}
%     \centering
%     \includegraphics[width=\linewidth]{arw}
%     \caption{Edge formation in \texttt{ARW}: consider
%     an incoming node $u$ with outdegree ${m=3}$ and attribute value {$B(u)=\textsc{red} \in \{\textsc{red},\textsc{green}\}$}.
%     In fig. 3a, $u$ joins the network and selects seed $v_a$ via \textsc{Select-Seed}.
%     Then, in fig. 3b, $u$ initiates a \textsc{Random-Walk} and traverses from $v_a$ to $v_b$ to $v_c$.
%     Finally, $u$ jumps back to its seed $v_a$ and restarts the walk, as shown in fig. 3c.
%     Node $u$ halts the random walk after linking to $v_a$, $v_c$ \& $v_d$.
%     }
%     \label{fig:randomwalk}
% 	% \vspace{-8pt}
% \end{figure*}

\section{Attributed Random Walk Model}
\label{sec:Proposed Model}
We propose an Attributed Random Walk (\texttt{ARW}) model to explain the emergence
of key structural properties of real-world networks through {entirely local}
edge formation mechanisms.

Consider a stylized example of how a researcher might go about finding relevant
papers to cite. First, the researcher broadly identifies one or more {relevant}
papers, possibly with the help of external information (e.g. Google
Scholar). These initial set of papers act as seed nodes.  Then, acting under
time and information constraints, she will examine papers cited by the seed and
papers that cite the seed. Thus, she navigates a chain of
backward and forward references to identify {similar}, relevant papers.
Next, through careful analysis, she will cite a subset of these papers. Similarly, users in
online social networks might form new friendships by navigating their social
circle (e.g., friends of friends) to find similar others.

\texttt{ARW} grows a directed network as new nodes join the network. The
mechanism is motivated by the stylized example: an incoming node selects a seed
node and initiates a random walk to explore the network by navigating through
neighborhoods of existing nodes. It halts the random walk after connecting to a
few visited nodes.

In this section, we describe the edge formation mechanisms underlying
\texttt{ARW}, explain how \texttt{ARW} unifies multiple sociological phenomena,
discuss model interpretability and summarize methods required to fit \texttt{ARW}
to network data.
% the observations from empirical data as well as Social Science studies. Then,
% we discuss the methods required to fit \texttt{ARW} to data.

\subsection{Model Description}
\label{sub:Model Description}
The Attributed Random Walk (\texttt{ARW}) model grows a directed network $\{\hat{G}_t\}^T_{t=1}$
in $T$ time steps.
More formally, at every discrete time step $t$, a
new node $u$, with attribute value $B(u)$, joins the network $\hat{G}_t$.
After joining the network, node $u$ forms $m(t)$ edges to
existing nodes.
% At time $t$, $G_t$ consists of ${|V_t|=|V_0|+t}$ nodes,
% ${|E_t|=|E_{t-1}|+m(t)}$ edges and the set of attribute values ${A_t = A_{t-1}
% \cup \{A(u)\}}$.
% The out-degree of incoming nodes increases over time to
% reflect the nonlinear growth and densification of real-world networks.
% We discuss the
% issue of initializing $G_0$, sampling attribute values of inomcing nodes and modeling
% densification in \Cref{sub:Model Fitting}.

The edge formation mechanism consists of two components: \textsc{Select-Seed} and
\textsc{Random-Walk}. As shown in~\Cref{fig:randomwalk}, an incoming node $u$ with attribute value $B(u)$ that joins the
network at time $t$ first selects a seed node using \textsc{Select-Seed}.

\textsc{Select-Seed} accounts for homophilic preferences of incoming nodes using
parameters $\psame$ and $\pdiff$ to tune attribute preferences.
In~\Cref{fig:randomwalk}, node $u$ selects a seed node and initiates a
random walk using \textsc{Random-Walk} to form $m(t)$ links:
\\\\
\tikzstyle{background rectangle}=[thin,draw=black]
\begin{tikzpicture}[show background rectangle]
	\node[align=left, text width=.93\linewidth, inner sep=.5em]{
		(1) With probability $\nicefrac{\psame}{\psame+\pdiff}$, randomly select a seed node
		from existing nodes that have the same attribute value, $B(u)$.

		\vspace{1mm}
		(2) Otherwise, with probability $\nicefrac{\pdiff}{\psame+\pdiff}$, randomly select a seed node from existing nodes that
		do \textit{not} have the same attribute value, $B(u)$.
	};
	\node[xshift=3ex, yshift=-.7ex, overlay, fill=white, draw=white, above
	right] at (current bounding box.north west) {
		\textsc{Select-Seed}
	};
\end{tikzpicture}

% The attribute parameter $p_a$ incorporates the attribute preferences of incoming nodes
% into the model.

The \textsc{Random-Walk} mechanism consists of four parameters: attribute-based parameters
$\psame$ \& $\pdiff$ model edge formation decisions and the jump parameter $\pjump$ \&
out-link parameter $\pout$ characterize random walk traversals:
\\\\
\tikzstyle{background rectangle}=[thin,draw=black]
\begin{tikzpicture}[show background rectangle]
	\node[align=left, text width=.93\linewidth, inner sep=.5em]{
		(1) At each step of the walk, new node $u$ visits node $v_i$.
		\begin{itemize}
			\item If $B(u)=B(v_i)$, $u$ links to $v_i$ with probability $\psame$
			\item Otherwise, $u$ links to $v_i$ with probability $\pdiff$
		\end{itemize}

		\vspace{1mm}
		(2) Then, with probability $\pjump$, $u$ jumps back to seed $s_u$.

		\vspace{1mm}
		(3) Otherwise, with probability ${1-\pjump}$, $u$ continues to walk.
		It picks an outgoing edge with prob. $\pout$ \textit{or}
		an incoming edge with prob. $1-\pout$ to visit a neighbor of $v_i$.

		\vspace{1mm}
		(4) Steps 1-3 are repeated until $u$ links to $m(t)$ nodes.
	};
	\node[xshift=3ex, yshift=-.7ex, overlay, fill=white, draw=white, above
	right] at (current bounding box.north west) {
		\textsc{Random-Walk}
	};
\end{tikzpicture}

When attribute data is absent, \texttt{ARW} simplifies further. A single link parameter $\plink$ replaces both attribute parameters $\psame$ \& $\pdiff$.
\textsc{Select-Seed} reduces to uniform seed selection and in
\textsc{Random-Walk}, the probability of linking to visited nodes equals $\plink$.
%  \textsc{Select-Seed} simplifies to selecting an existing
% node uniformly at random and \textsc{Random-walk} simplifies to selecting an existing
% node uniformly at random.\texttt{ARW} only requires a single link
% parameter $\plink$, instead of both $\psame$ \& $\pdiff$. Additionally,

Note that \texttt{ARW} has two exogenous parameters: the out-degree
$m(t)$ and attribute $B(u)$ of incoming nodes.
The attribute distribution varies with time as new attribute values (e.g., journals) crop up, necessitating an exogenous
parameter. The parameter $m(t)$ is the mean-field value
of out-degree $m$ at time $t$ in the observed network.
While it is straightforward to model $m(t)$
endogenously by incorporating a densification power-law \texttt{DPL} exponent,
exogenous factors (e.g., venue, topic) may influence node out-degree.
%  similar to the parameter $m$ in the classic Preferential-Attachment
% model~\cite{barabasi1999emergence}, except that $m(t)$ is the mean-field value
% of out-degree $m$ at time $t$ in the observed network.
% While it is straightforward to model $m(t)$
% endogenously by incorporating a densification power-law \texttt{DPL} exponent
% to \texttt{ARW}, we decided against it, since exogenous factors may
% also explain changes to $m(t)$. For example, conference venue and paper topic can
% influence the number of citations in a paper. Moreover, our analysis indicates
% that papers that join networks earlier tend to have fewer citations on average,
% perhaps explained by availability of {fewer} papers to cite.

Next, we explain how each parameter is necessary to conform to normative
behavior of individuals in evolving networks.

\subsection{\texttt{ARW} and Normative Behavior}
\label{sub:Model Interpretation}
The Attributed Random Walk model unifies multiple sociological phenomena
into its edge formation mechanisms.

\newtheoremstyle{exampstyle}
  {2pt} % Space above
  {2pt} % Space below
  {\itshape} % Body font
  {} % Indent amount
  {\bfseries} % Theorem head font
  {.} % Punctuation after theorem head
  {.25em} % Space after theorem head
  {} % Theorem head spec (can be left empty, meaning `normal')

\theoremstyle{exampstyle} \newtheorem{ph}{Phenomenon}

\begin{ph}
	(Limited Resources) Individuals are boundedly rational~\cite{simon1972theories,gigerenzer1996reasoning,lipman1995information}
	actors that form edges under constraints of limited information and partial network access.
\end{ph}
% \texttt{ARW} uses random walk traversals to incorporate constraints of limited information
% and partial network access.
As shown in ~\Cref{fig:randomwalk}, \textsc{Random-Walk} only requires information only about the
1-hop neighborhood of a few visited nodes, thereby accounting for the constraints of limited information and partial network access.
% A new node $u$ selects a seed node from which it
% initiates a biased random walk. Then, $u$ uses simple rules to connect to each visited
% nodes probabilistically and halts the walk after forming $m(t)$ edges, as shown in~\Cref{fig:randomwalk}. Random walks require information only about the
% 1-hop neighborhood of a few visited nodes, thereby accounting for  the constraints of limited information and partial network access.

\begin{ph}
	(Structural Constraints) Network distance
	act as a constraint that limits long-range connections.  \cite{35626}
\end{ph}

We incorporate structural constraints using $\pjump$, the probability with
which a new node jumps back to its seed node after every step of the random walk. This implies
that the probability with which the new node is at most $k$ steps from its seed node is $(1-\pjump)^k$;
as a result, $\pjump$ controls the extent to which nodes' random walks explore the network to form edges.

\begin{ph}
	(Triadic Closure) Nodes with common neighbors have an
	increased likelihood of forming a connection. \cite{simmel1950sociology}
\end{ph}

When attribute data is absent, \texttt{ARW} controls
the effect of triadic closure on link formation using $\plink$ because
with probability proportional to $\plink^2$,
a new node $u$ closes a triad through its random traversal by linking to both, a visited node
and its neighbor,
% Similarly, in attributed networks, the probability of triad completion equals $pq$,
% where $p$ and $q$ can equal $\psame$ or $\pdiff$, depending on the attribute values of
% $u$ and the visited nodes.

\begin{ph}
	(Attribute Homophily) Nodes that have similar attributes are more likely
	to form a connection. \cite{mcpherson2001birds}
\end{ph}
The attribute parameters $\psame$ and $\pdiff$ modulate
attribute assortativity. When $\psame > \pdiff$, nodes are more likely to connect if they share
the same attribute value, thereby resulting in a homophilic network over time. Similarly,
$\psame < \pdiff$ and $\psame=\pdiff$ make edge formation heterophilic and attribute agnostic respectively.

\begin{ph}
	(Preferential Attachment) Nodes tend to link to high degree nodes that have more
	visibility. \cite{barabasi1999emergence}
\end{ph}
\texttt{ARW} controls preferential attachment by adding structural bias to the
random walk traversal using outlink parameter $\pout$, instead of relying on the
global degree distribution. Random walks that traverse outgoing edges only
(i.e., $\pout =1$) eventually visit old nodes that tend to have high in-degree.
Similarly, random walks that traverse incoming edges only (i.e., $\pout=0$) visit
recently joined nodes that tend to have low indegree. As a result, we use
$\pout$ to adjust bias towards node degree.
% effect of preferential attachment on edge formation.

To summarize: \texttt{ARW} incorporates five well-known sociological
phenomena---limited resources; structural constraints; triadic closure;
attribute homophily; preferential attachment---into a single edge formation
mechanism based on random walks.

\subsection{Model Interpretability}
\texttt{ARW} parameters intuitively shape key structural properties:
in-degree distribution, local clustering, path length and attribute assortativity.

In order to understand how global network properties vary as functions of
\texttt{ARW} parameters, we explore the parameter space of the model. As
described in~\Cref{sub:Model Description}, \texttt{ARW} uses two
parameterizations to model networks with or without attribute data. We analyze
network structure and attribute assortativity using $(\plink, \pjump, \pout)$
and $(\psame, \pdiff, \pjump, \pout)$ respectively.

\begin{figure}[b]
 % \vspace{-10pt}
 \centering
 \includegraphics[width=\columnwidth]{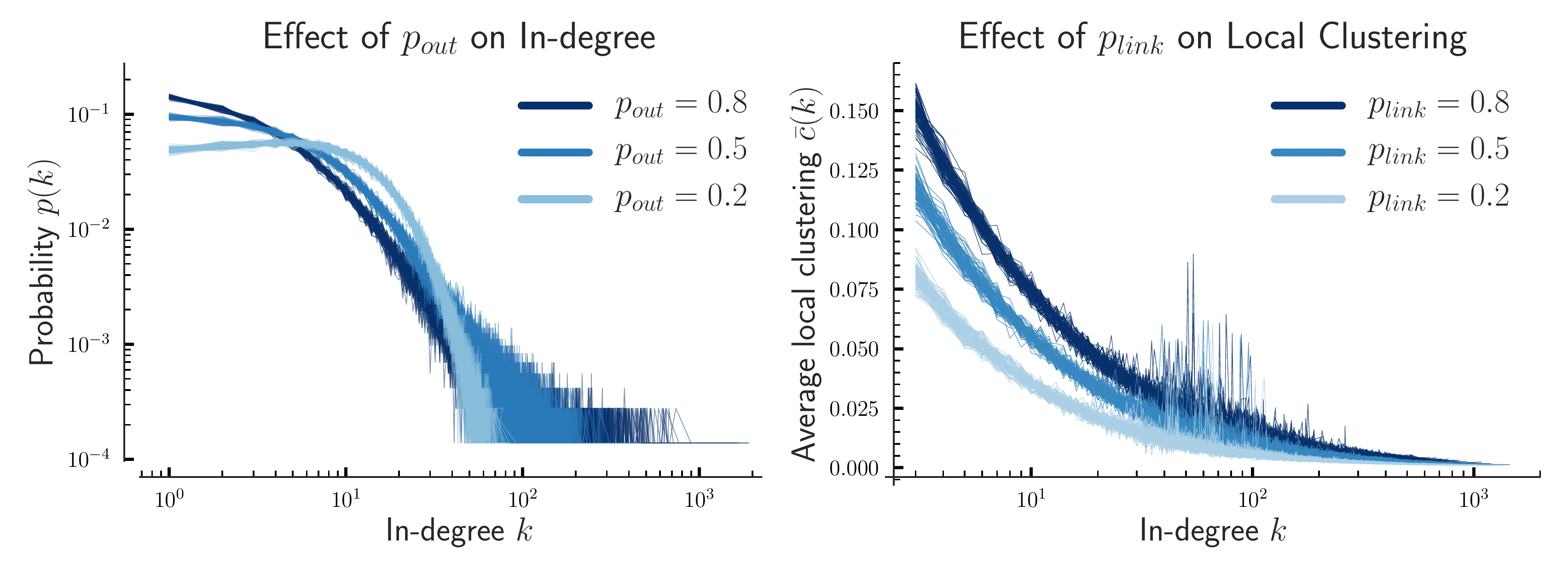}
 \caption
 {Effect of $\pout$ and $\plink$ on in-degree and local clustering.
 The left and right subplots show how increasing $\pout$ and $\plink$ yield
 in-degree distributions with heavier tails and neighbors with higher clustering
 by adding structural bias towards well-connected and proximate nodes respectively.}
 \label{fig:inter1}
 % \vspace{-10pt}
\end{figure}

\Cref{fig:inter1} illustrates how in-degree and local clustering depend on $\pout$ and $\plink$.
Increasing $\pout$ steers random walks towards older nodes that tend to have higher in-degree.
Over time, as more nodes join the network, initial differences in degree amplify, resulting in heavy-tailed distributions.
In ~\Cref{fig:inter1}, we observe that increasing $\pout$ from $0.2$ to $0.8$ shifts probability
mass from average degree nodes (\texttt{B}) to hubs (\texttt{C}) and low degree nodes (\texttt{A}).
As a result, $\pout$ controls the extent to which hubs skew the in-degree distribution.
Similarly, local clustering increases as a function of $\plink$ because $\plink$ implicitly
controls the rate at which new nodes close triads by linking to adjacent nodes in their random walks.

\begin{figure}
 % \vspace{-10pt}
 \centering
 \includegraphics[width=1.05\columnwidth]{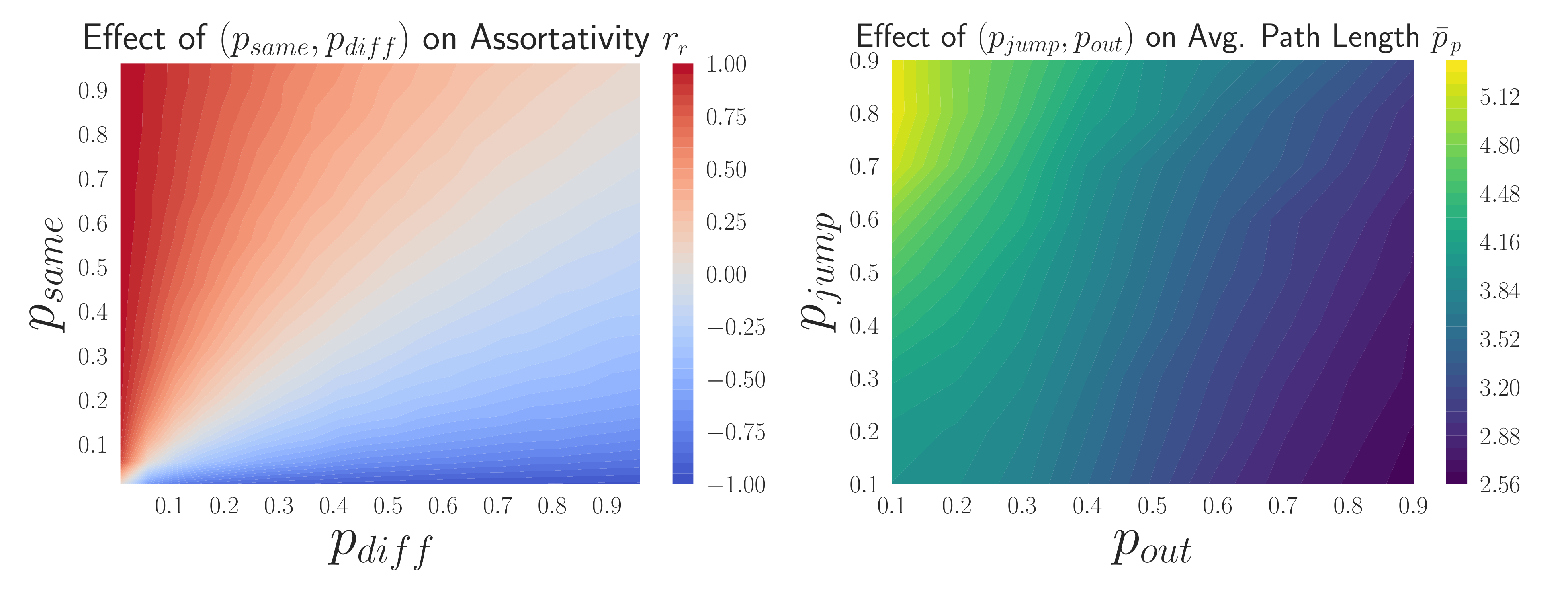}
 \caption{
 	Effect of $(\psame, \pdiff)$ and $(\plink, \pjump)$ on attribute assortativity
	and average path length.
	In the left, we observe that
 	increasing $\psame-\pdiff$ leads to higher assortativity by making edge formation more homophilic.
	Similarly, decreasing $\pjump$ while increasing $\pout$ shortens average path length
 	by enabling incoming nodes to form connections in disparate regions of the network.
 }
 \label{fig:inter2}
 % \vspace{-17pt}
\end{figure}

We use contour plots to visualize how $(\pdiff, \psame)$
and $(\plink, \pjump)$ alter attribute assortativity and average path length.
As shown in the left subplot of~\Cref{fig:inter2}, $\psame-\pdiff$ tunes the extent to which
attributes influence edge formation. Increasing $\psame-\pdiff$ increases
attribute assortativity by amplifying nodes' propensity to link to similar
nodes, which subsequently increases the fraction of edges between similar
nodes. More importantly, when $\psame-\pdiff$ remains constant, increasing
$(\psame, \pdiff)$ raises local clustering without altering attribute
assortativity. In the right subplot, we observe that increasing $\pout$ while
decreasing $\pjump$ shortens the average path length. This is because low
values of $\pjump$ do not restrict incoming nodes to the local neighborhood
of their seed nodes,
thereby allowing incoming nodes to visit and form edges to nodes that are
structurally distant. Additionally, increasing $\pout$ results in greater number of hubs,
which in turn act as intermediate nodes to connect nodes via short path
lengths.

Thus,
\texttt{ARW} unifies multiple sociological phenomena at the local level
as well as intuitively controls key global network properties.

\subsection{Model Fitting}
\label{sub:Model Fitting}

We now briefly describe methods to estimate model parameters,
initialize $\hat{G}$, densify $\hat{G}$ over time and sample nodes' attribute values.

\textit{Parameter Estimation}. We use grid search to estimate
the four parameters using evaluation metrics and selection criterion described in~\Cref{sub:Experimental Setup}.
% The parameter estimation task consists of finding the set of parameters values
% for $(\psame, \pdiff, \pjump, \pout)$ that best preserve the structural properties of an
% observed network $G$.
% Note that other derivative-free optimization methods such as the Nelder-Mead~\cite{nelder1965simplex} method
% can be used to speed-up parameter estimation.

\textit{Initialization}. \texttt{ARW} is
sensitive to a large number of weakly connected components (\texttt{WCC}s) in
initial network $\hat{G}_0$ because incoming nodes only form edges to nodes
in the same \texttt{WCC}. To ensure that $\hat{G}_0$ is weakly
connected, we perform an undirected breadth-first search on the observed,
to-be-fitted network $G$ that starts from the oldest node and halts after
visiting $0.1\%$ of the nodes. The initial network $\hat{G}_0$ is the small \texttt{WCC}
induced from the set of visited nodes.
% Simpler initialization methods
% such as sampling $\hat{G}_0$ from the Erdos-Renyi model or Watts-Strogatz model
% yield similar results.

\textit{Node Out-degree}.
Node out-degree increases non-linearly over time in real-world networks.
We coarsely mirror the growth rate of observed network $G$ as follows.
Each incoming node $u$ that joins $\hat{G}$ at time $t$ corresponds to some
node that joins the observed network $G$ in year $y(t)$; the number of edges $m(t)$
that $u$ forms is equal to the average out-degree of nodes that join $G$ in year $y(t)$.

\textit{Sampling Attribute Values}.
The distribution over nodal attribute values $P_{\textsc{g}}(B)$ tends to change over time.
% For instance, the attribute distribution over journals in the \texttt{APS} citation
% network changes over time as old journals decay in popularity and new journals gain traction.
The change in the attribute distribution over time is an exogenous factor and varies for every network.
Therefore, we sample the attribute value $B(u)$ of node $u$, that joins $\hat{G}$ at time
$t$, from $P_{\textsc{g}}(B{\mbox{ | year}=y(t)})$, the observed attribute
distribution conditioned on the year of arrival of node $u$.

To summarize, \texttt{ARW}
intuitively describes how individuals form edges under resource constraints.
\texttt{ARW} uses four parameters ---$\psame$, $\pdiff$, $\pjump$, $\pout$--- to incorporate
individuals' biases towards similar, proximate and high degree nodes.
% We note that a mathematical analysis of \texttt{ARW} appears to lead to novel
% theoretical questions about random walks on directed graphs.
% Unfortunately, unlike the case of undirected graphs, there is no
% analytical closed form expression for stationary distributions of random walks
% on general directed graphs, thereby making the analysis of random walk models
% like \texttt{ARW} quite complex. \harshay{Improve theory justification
Next, we discuss our experiments on the performance of $\texttt{ARW}$
in accurately preserving {multiple} structural and attribute properties of real networks.

% \clearpage
%!TEX root = ../draft.tex

\begin{figure*}[t]
	\vspace{-15pt}
	\centering
	\makebox[\textwidth][c]{\includegraphics[width=.97\textwidth]{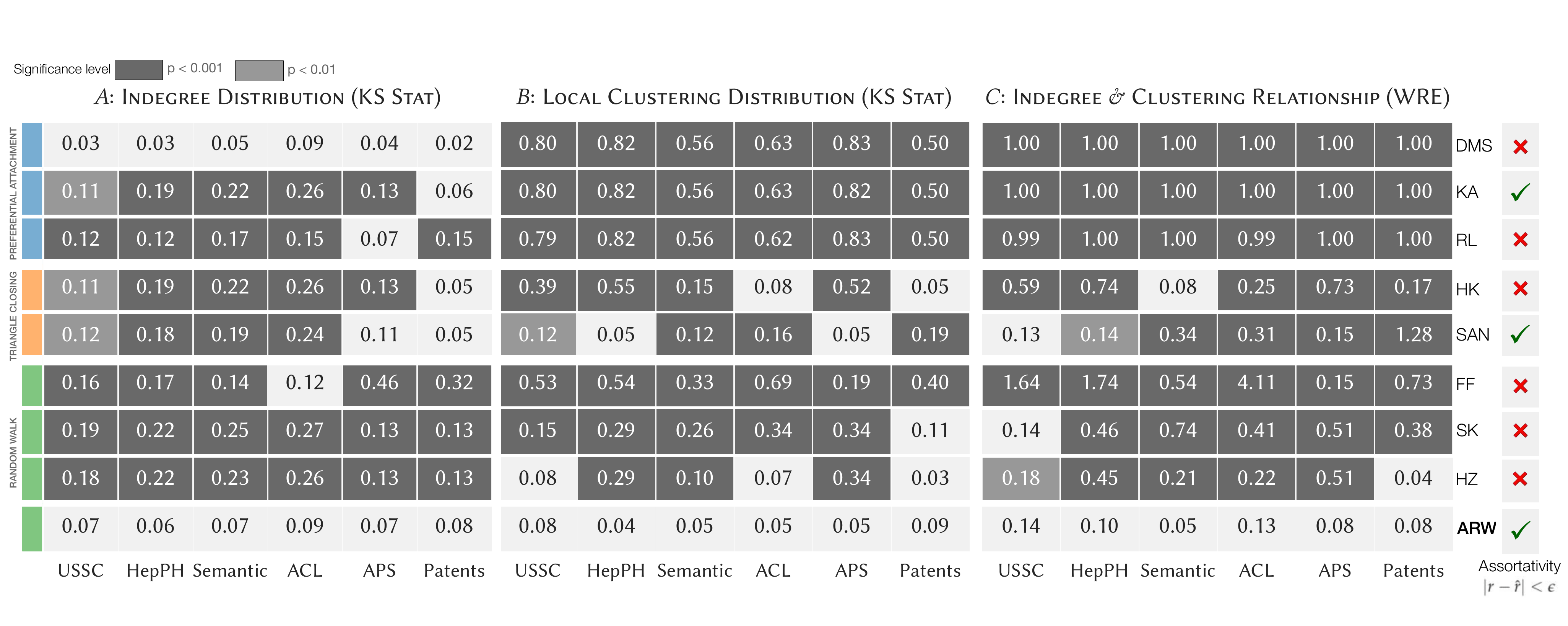}}
	\vspace{-16pt}
	\caption{
		Modeling network structure. We assess the extent to which network models
		fit key structural properties of six real-world networks. Tables 5A, 5B and 5C
		measure the accuracy of eight models in fitting the in-degree distribution,
		local clustering distribution, in-degree \& clustering relationship
		respectively and global attribute assortativity.
		Existing models tend to underperform because they either disregard
		the effect of factors such as triadic closure and/or homophily
		or are unable to generate networks with varying structural properties.
		Our model, \texttt{ARW}, jointly preserves all three properties accurately and often
		performs considerably better than existing models:
		the cells are shaded gray or dark gray if the proposed model \texttt{ARW} performs
		better at significance level $\alpha=0.01$ ( \lightgraybg{ }) or $\alpha=0.001$ ( \darkgraybg{ })
		respectively.
	}
	\vspace{-10pt}
	\label{fig:exp_table}
\end{figure*}

\section{Modeling Network Structure}
\label{sec:Experiments}
In this section, we evaluate \texttt{ARW}'s performance in preserving
real-world network structure relative to well-known growth models.
% In~\Cref{sub:Experimental Setup}, we begin by describing existing growth models and the evaluation metrics
% used in the experiments. Then, we discuss our results.

\subsection{Setup}
\label{sub:Experimental Setup}

% We first briefly summarize the existing models used in the experiments.
In this subsection, we introduce eight representative growth models
and describe evaluation metrics used to fit models to the datasets.

\textit{State-of-the-art Growth Models}. We compare \texttt{ARW} to eight state-of-the-art
growth models representative of the key edge formation
mechanisms: preferential attachment, fitness, triangle closing and random walks.
Two of the eight models account for attribute homophily and preserve attribute mixing patterns,
as listed below:

		\noindent{\textbf{(1) Dorogovtsev-Mendes-Samukhin model}} \cite{dorogovtsev2000structure}  (\texttt{DMS})
		is a preferential attachment model that generates directed scale-free graphs. In this model,
		the probability of linking to a node is proportional to the sum of its in-degree and ``initial attractiveness.''

		\noindent{\textbf{(2) Relay Linking model}} \cite{singh2017relay} (\texttt{RL}) comprises
		preferential attachment models for directed networks that use relay linking to model
		node popularity over time. We use the iterated preferential relay-cite (IPRC) variant, which best fits
		real-world network properties.

		\noindent{\textbf{(3) Kim-Altmann model}} \cite{kim2017effect} (\texttt{KA}) is a fitness-based model that defines
		fitness as the product of degree and attribute similarity. It generates {attributed} networks
		with assortative mixing and power law degree distribution.
		To generate directed networks, we modify \texttt{KA} to form directed edges to nodes in proportion to their in-degree.

		\noindent{\textbf{(4) Holme-Kim model}} \cite{holme2002growing} (\texttt{HK}) is a preferential attachment model
		that generates scale-free, clustered, undirected networks using a triangle-closing mechanism.
		To generate directed networks, we modify \texttt{HK} to form directed edges to nodes in proportion to their in-degree
		and close triangles in their undirected 1-hop neighborhood.

		\noindent{\textbf{(5) Social Attribute Network model}} \cite{gong2012evolution} (\texttt{SAN}) generates
		scale-free, clustered, attributed networks via attribute-augmented
		preferential attachment and triangle closing processes.
		We modify \texttt{SAN} to create directed edges and thereby produce directed networks.
		% We also note that
		% the edge formation mechanism in \texttt{SAN} considerably simplifies for bibliographic network datasets,
		% wherein all edges are formed at once.

		\noindent{\textbf{(6) Herera-Zufiria model}} \cite{saramaki2004scale} (\texttt{SK})
		is a random walk model that generates scale-free, undirected networks with tunable average clustering.
		In order to generate directed networks, we allow the random walk mechanism in \texttt{SK} to traverse edges in any direction.

		\noindent{\textbf{(7) Saramaki-Kaski}} \cite{herrera2011generating} (\texttt{HZ}) is a random walk model
		that generates scale-free networks with tunable average local clustering. To generate directed networks,
		we modify \texttt{HZ} to allow its random walk mechanism to traverse edges in any direction.

		\noindent{\textbf{(8) Forest Fire model}} \cite{leskovec2005graphs} (\texttt{FF}) is a recursive random walk model
		that can generate directed networks with shrinking diameter over time,
		heavy-tailed degree distributions and high clustering.

\textit{Ensuring Fair Comparison}. To ensure fair comparison, we modify existing models in three ways.
First, for \texttt{DMS}, \texttt{SAN}, \texttt{KA} do not have an explicitly defined initial graph,
so we use initialization method used for \texttt{ARW}, described in~\cref{sub:Model Fitting}. Second, we extend
models that use constant node outdegree $m$ by increasing outdegree over time $m(t)$
using the method described in~\cref{sub:Model Fitting}. In the absence of model-specific parameter estimation methods,
we use grid search to estimate the parameters of every network model, including \texttt{ARW},
using evaluation metrics and selection criterion described below.

\textit{Evaluation Metrics}.
We evaluate the network model fit by comparing four structural properties of ${G}$ \& $\hat{G}$:
degree distribution, local clustering distribution, degree-clustering relationship
and attribute assortativity. We use Kolmogorov-Smirnov (\texttt{KS}) statistic to compare in-degree
\& local clustering distributions. We compare the degree-clustering relationship in $G$ and $\hat{G}$ using
Weighted Relative Error (\texttt{WRE}), which aggregates the relative error
between the average local clustering $c(k)$ and $\hat{c}(k)$ of nodes with in-degree $k$
in $G$ and $\hat{G}$ respectively; The relative error between $c(k)$ and $\hat{c}(k)$
is weighted in proportion to the number of nodes with in-degree $k$ in $G$.

Jointly preserving multiple structural properties is a multi-objective optimization
problem; model parameters that accurately preserve the degree distribution
(i.e. low \texttt{KS} statistic) may not preserve the clustering distribution.
Therefore, for each model, the selection criterion for the grid search parameter estimation method
chooses the model parameters that minimizes the $\ell^2$-norm of the aforementioned evaluation metrics.
Since the metrics have different scales, we normalize the metrics before computing the $\ell^2$-norm
to prevent unwanted bias towards any particular metric.
We note that the parameter sensitivity of the Forest Fire (\texttt{FF}) model necessitates
a manually guided grid search method.

\begin{figure*}
	\centering
	\makebox[\textwidth][c]{\includegraphics[width=\textwidth]{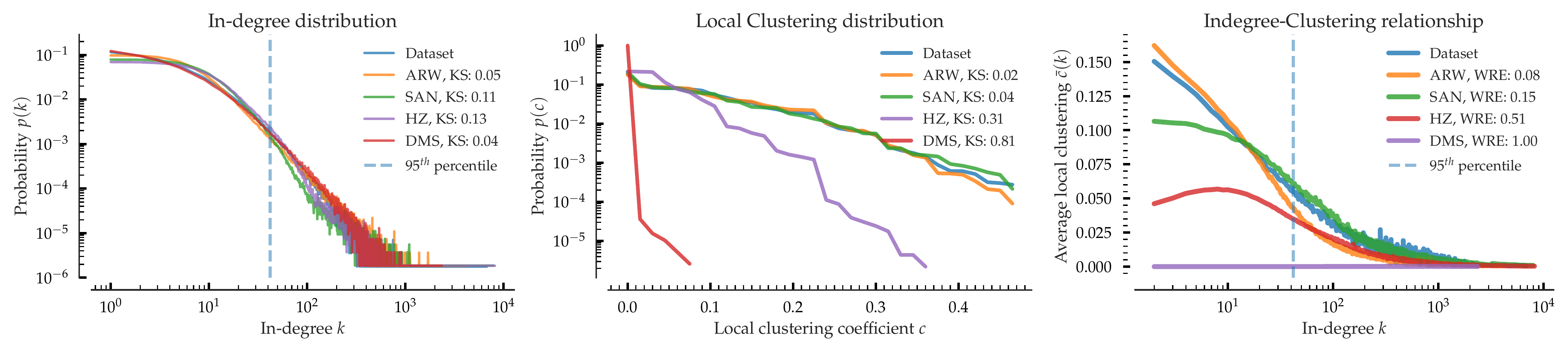}}
	\caption{
		Performance of \texttt{ARW} in accurately preserving key global structural properties
		of the \texttt{APS} network dataset relative to state-of-the-art, representative
		network models. Existing models such as \texttt{DMS} and \texttt{HK} cannot preserve high
		local clustering.
		% Although \texttt{SAN} preserves the univariate in-degree
		% and local clustering distributions, it does not account for the correlation between in-degree and clustering.
		Moreover, the triangle closing mechanism in \texttt{SAN} incurs high Weighted Relative Error (\texttt{WRE}) because
		it cannot explain why low in-degree nodes have high local clustering. \texttt{ARW} outperforms existing network models
		in jointly preserving all three structural properties, in addition to attribute mixing patterns.}
	\label{fig:aps_fits}
\end{figure*}

% \vspace{-11pt}
\subsection{Results}
\label{sub:Experimental Results}

Now, we evaluate the performance of \texttt{ARW} relative to eight well-known
existing models on the datasets introduced in~\Cref{sec:Datasets}.
\Cref{fig:exp_table} tabulates the evaluation metrics for every pair of model
and dataset. These metrics measure the accuracy with which the fitted models
preserve key global network properties: degree distribution, local clustering distribution,
and indegree-clustering relationship.
% We do not compare the extent to which these models preserve attribute
% assortativity because the attribute related model parameters can be
% independently tuned up to arbitrary precision. Instead,

To evaluate the performance of these models, we first fit each model
to all network datasets $G$ in~\Cref{sec:Datasets}.
Thereafter, we compare the structural properties of network dataset $G$ and network $\hat{G}$
generated by the fitted model using evaluation metrics in~\Cref{sub:Experimental Setup}. We average out
fluctuations in $\hat{G}$ over 100 runs.
 % data from these runs also help us conduct statistical tests.

We use one-sided permutation tests \cite{good2013permutation} to evaluate the relative
performance of \texttt{ARW}. If \texttt{ARW} performs better than a model on a dataset
with significance level $\alpha=0.01$ or $\alpha=0.001$, the corresponding cells in~\Cref{fig:exp_table}
are shaded gray ( \lightgraybg{ }) or dark gray (~\darkgraybg{ }) respectively.
We also group models that have similar edge formation mechanisms by color-coding the
corresponding rows in~\Cref{fig:exp_table}.  We use green ticks in~\Cref{fig:exp_table} to
annotate models that preserve assortativity up to two decimal places.
% The performance of a model depends on the effectiveness of its underlying
% edge formation mechanisms. Therefore,  we group models (i.e. color-coded rows in table \ref{fig:exp_table}) based on their
% underlying edge formation mechanism: preferential attachment (blue), preferential attachment with
% triangle closing (orange) and random walk (green) mechanisms.

~\Cref{fig:exp_table} shows that existing models fail to jointly preserve
{multiple} structural properties in an accurate manner. This is because existing
models either disregard important mechanisms such as triadic closure and homophily
or are not flexible enough to generate networks with varying structural properties.
% For instance, the preferential attachment model \texttt{DMS}
% can accurately fit the heavy-tailed degree distributions but does not
% account for local clustering.
% On the other hand, the attributed network model \texttt{SAN} tries to preserve
% all four properties but cannot do so accurately.

\textbf{Preferential attachment models}: \texttt{DMS}, \texttt{RL}
and \texttt{KA} preserve in-degree distributions but disregard
clustering. \texttt{DMS} outperforms other models in accurately modeling
degree distribution (\Cref{fig:exp_table}A) because its ``initial attractiveness''
parameter can be tuned to adjust preference towards low degree nodes. Unlike \texttt{KA}, however,
\texttt{DMS} cannot preserve global assortativity.
% \texttt{KA} outperforms \texttt{DMS} in preserving global assortativity because \texttt{KA}
% uses an attribute similarity parameter to model attribute mixing patterns.
However, by assuming that successive edge formations are independent, both models disregard
triadic closure and local clustering. (\Cref{fig:exp_table}B \&~\Cref{fig:exp_table}C).

\textbf{Triangle Closing Models}: \texttt{HK} and \texttt{SAN} are preferential attachment models
that use triangle closing mechanisms to generate scale-free networks with high average
local clustering.
% Note that \texttt{HK} and \texttt{KA} fit degree distributions with the same \texttt{KS} statistic
% (\Cref{fig:exp_table}A) because they lack parameters that can generate varying degree distributions.
While triangle closing leads to considerable improvement over \texttt{DMS}
and \texttt{KA} in modeling local clustering, \texttt{HK} and \texttt{SAN} are not flexible enough
to preserve local clustering in {all} datasets (see~\Cref{fig:exp_table}B \&~\Cref{fig:exp_table}C).
%  Nevertheless,
% barring one or two datasets in tables \ref{fig:exp_table}B and \ref{fig:exp_table}C,
% these models cannot accurately preserve the local clustering distribution and in-degree-clustering
% relationship observed in real networks.

\textbf{Existing random walk models}: \texttt{FF}, \texttt{SK}, and \texttt{HZ}
cannot accurately preserve structural properties of real-world network datasets.
The recursive approach in \texttt{FF} considerably overestimates local clustering.
because nodes perform a probabilistic breadth-first search and link to \textit{all} visited/burned
nodes.
\texttt{SK} and \texttt{HZ} can control local clustering to some extent, as
nodes perform a single random walk and link to each visited node with tunable probability $\mu$.
However, both models lack control over the in-degree distribution. Furthermore, existing random walk models
disregard attribute homophily and do not account for attribute mixing patterns.

\begin{figure}[b]
	\centering
	% \vspace{-3pt}
	\includegraphics[width=.95\linewidth]{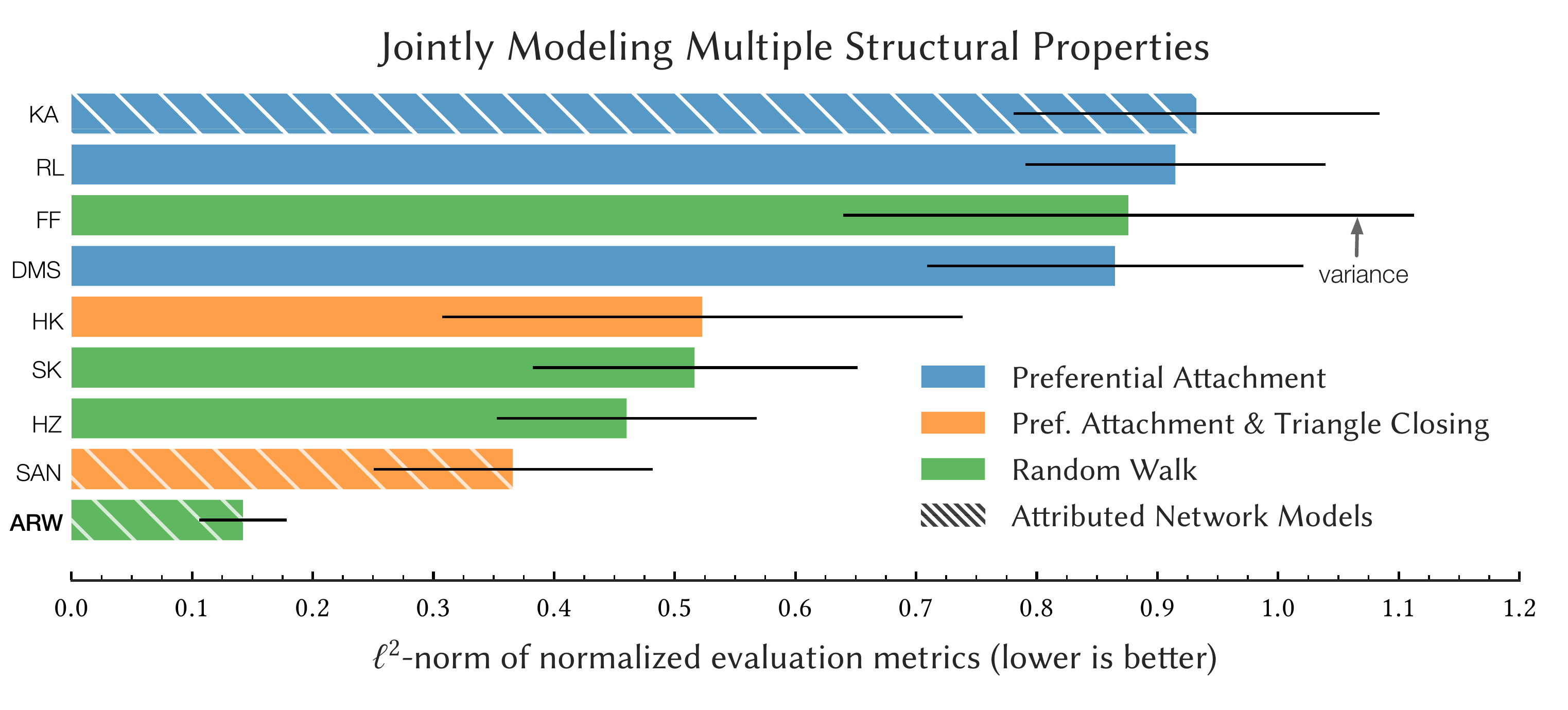}
	% \vspace{-10pt}
	\label{fig:barplot}
		\caption{\texttt{ARW} outperforms
			existing network models in jointly preserving key structural properties---in-degree
			distribution, local clustering distribution and degree-clustering relationship---
			by a significant margin of 2.5x-10x.
		}
\end{figure}

\textbf{Attributed Random Walk model}:~\Cref{fig:exp_table} clearly indicates the effectiveness
of \texttt{ARW} in {jointly} preserving multiple
global network properties. \texttt{ARW} can generate networks with tunable
in-degree distribution by adjusting nodes' bias towards high degree nodes
using $\pout$. As a result, \texttt{ARW} accurately preserves
in-degree distributions (\Cref{fig:exp_table}A), often significantly better
than all models except \texttt{DMS}. Similarly, \texttt{ARW} matches the local clustering
distribution  (\Cref{fig:exp_table}B) and in-degree \& clustering relationship
(\Cref{fig:exp_table}C) with high accuracy using $\pjump$ and
$\plink$. Similarly, \texttt{ARW} preserves attribute assortativity using
the attribute parameters $\psame$ and $\pdiff$.
Barring one to two datasets, \texttt{ARW} preserves all three properties significantly
better ($\alpha < 0.001$) than existing random walk models.

To summarize, \texttt{ARW} unifies five sociological phenomena
into a single mechanism to jointly preserve real-world network structure.
\section{Modeling Local Mixing Patterns}
\label{subsec:LocalMixing}

The global assortativity coefficient quantifies
the average propensity of links between similar nodes.
However, global assortativity is not a representative summary statistic of
heterogeneous mixing patterns observed in large-scale networks~\cite{peel2018multiscale}.
Furthermore, it does not quantify anomalous mixing patterns and fails to measure how mixing varies across a network.

We use local assortativity~\cite{peel2018multiscale} to measure varying
mixing patterns in an attributed network $G=(V,E,B)$ with attribute values $B=\{b_1...b_k\}$.
Unlike global assortativity that counts all edges between similar nodes, local assortativity
of node $i$, $\rlocal(i)$, captures attribute mixing patterns in the neighborhood of node
$i$ using a proximity-biased weight distribution $w_i$. The distribution
$w_i$ reweighs edges between similar nodes based on proximity to
node $i$. As~\citet{peel2018multiscale} indicate, there are multiple ways
to define node $i$'s weight distribution $w_i$ other than the prescribed
personalized pagerank weight distribution, which is prohibitively expensive to compute
for all nodes in large graphs.
We define $w_i$ as a uniform distribution over $N_2(i)$, the two-hop local neighborhood
of node $i$, to allow for a highly efficient
local assortativity calculation.
Intuitively, $\rlocal(i)$ compares the observed fraction of edges between similar nodes
in the local neighborhood of node $i$ to the expected fraction
if the edges are randomly rewired.
% % EQUATION START
% More formally, the local assortativity coefficient $\rlocal(i)$ of node $i$, with outdegree $m(i)$ and
% attribute value $b(i)$ is defined as follows:
% \begin{align*}
% 	\scriptsize \rlocal(i) = \frac{\overbrace{\frac{1}{|N(i)|}\sum\limits_{j \in N(i)}^{m(j) > 0} \sum_{k \in V} \frac{\mathcal{I}\{(j,k) \in E \wedge b(j)=b(k)\}}{m(i)} }^{\texttt{observed}}-\overbrace{\sum_{b \in B} e_{b}\cdot e_{b}}^{\texttt{random}}}{\underbrace{1}_{\max(\texttt{observed})}-\underbrace{\sum_{b \in B} e_{b} \cdot e_{b}}_\texttt{random}}
% \end{align*}
% % EQUATION END

\begin{figure}
	\centering
	% \vspace{-9pt}
	\includegraphics[width=1.05\columnwidth]{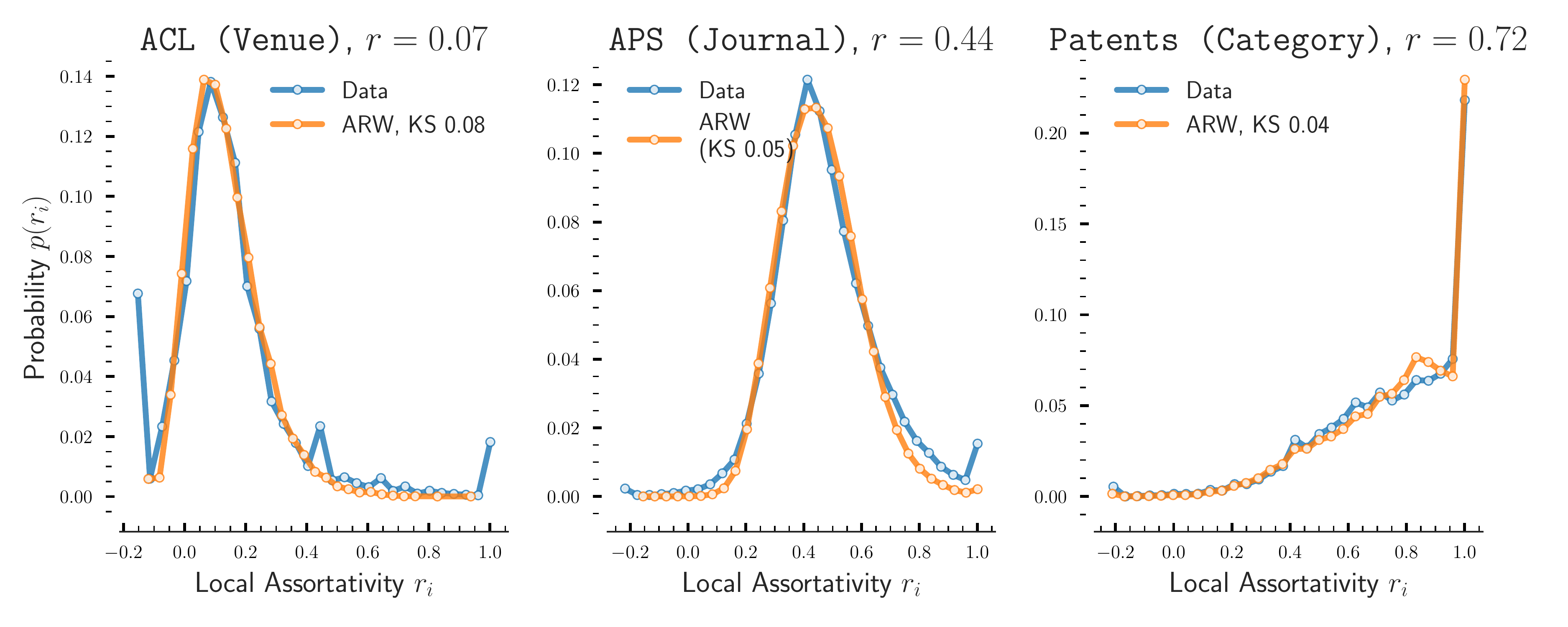}
	\caption{Local assortativity distributions of attributed networks \texttt{ACL}, \texttt{APS}
		and \texttt{Patents} reveal anomalous, skewed and heterophilic local mixing patterns.
		\texttt{ARW} accurately preserves local assortativity, but does not account for anomalous mixing patterns.}
	\label{fig:local_atty}
	\vspace{-8pt}
\end{figure}

As shown in~\Cref{fig:local_atty}, local assortativity distributions
of \texttt{ACL}, \texttt{APS} and \texttt{Patents} reveal anomalous, skewed
and heterophilic local mixing patterns that are not inferred via global assortativity.
Our model \texttt{ARW} can preserve
diverse local assortativity distributions with high accuracy even though nodes
share the same attribute parameters $\psame$ and $\pdiff$. This is because, in addition
to sampling attributes conditioned on time, \texttt{ARW}
incorporates multiple sources of stochasticity through its edge formation
mechanism. As a result, incoming nodes with fixed homophilic preferences can position
themselves in neighborhoods with variable local assortativity by (a) selecting a seed node in a region
with too few (or too many) similar nodes or (b) exhausting all its links before
visiting similar (or dissimilar) nodes.
We note that \texttt{ARW} is not expressive enough to model anomalous
mixing patterns; richer mechanisms such as sampling $\psame$ or $\pdiff$
from a mixture of Bernoullis are necessary to account for anomalous mixing patterns.

% \clearpage
%!TEX root = ../draft.tex

\section{Discussion}
\label{sec:Discussion}
In this section, we discuss weaknesses of triangle closing mechanisms,
the effect of out-degree on network diameter and limitations \& potential
modifications of our model \texttt{ARW}.

\subsection{Dissecting the Triangle Closing Mechanism}
\label{ss:tc}

\begin{figure}[b]
    \centering
    \includegraphics[width=.8\columnwidth]{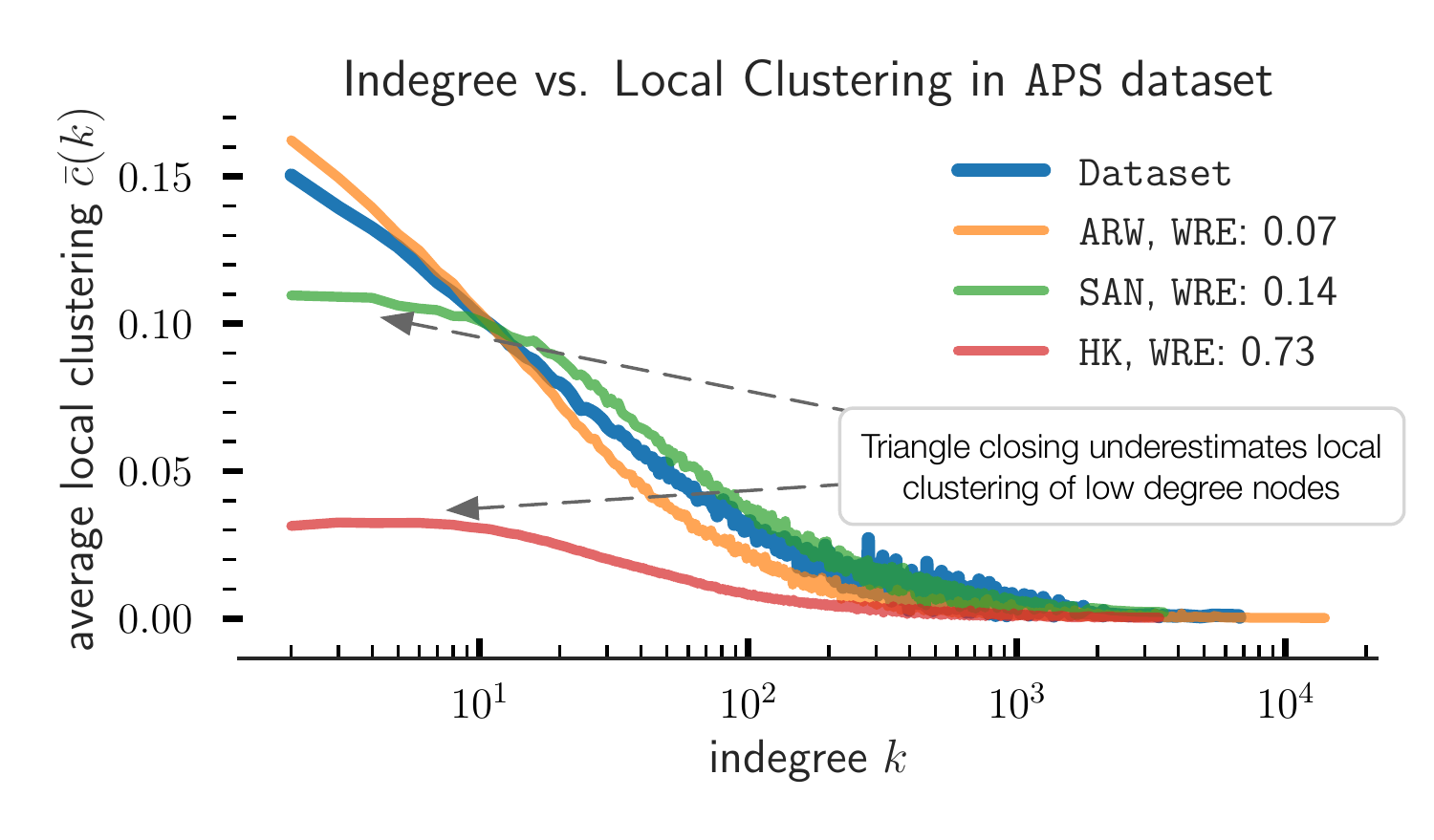}
    \caption{Triangle closing mechanisms used in \texttt{SAN} \texttt{HK} fail to
    model average local clustering of low in-degree nodes. In contrast,
    to accurately preserve local clustering, \texttt{ARW} uses {random walks} to visit
    low in-degree nodes and close triangles in their neighborhoods .}
    \label{fig:triangle_closing}
\end{figure}

A set of network models (e.g., \texttt{SAN} \cite{gong2012evolution} \& \texttt{HK} \cite{holme2002growing})
use triangle closing mechanisms to generate networks with
varying average local clustering. However, our experimental results
in~\Cref{sub:Experimental Results} show that models that rely on triangle closing
cannot explain local clustering distribution or bivariate degree-clustering
relationship accurately. To understand why, we examine the degree-clustering
relationship in the \texttt{APS} network in~\Cref{fig:triangle_closing}.

\Cref{fig:triangle_closing} reveals that models based on triangle closing mechanisms,
\texttt{SAN} and \texttt{HK}, considerably underestimate the local clustering of
nodes that have low in-degree. This is because incoming nodes in \texttt{SAN} and \texttt{HK}
tend to close triangles in the neighborhood of high in-degree nodes to which they
connect via preferential attachment. Local clustering plateaus as in-degree decreases because
triangle closing along with preferential attachment fail to form connections in neighborhoods
of low in-degree nodes. In contrast, \texttt{ARW} accurately models the degree-clustering relationship
because incoming nodes initiate random walks and close triangles in neighborhoods of low in-degree
seed nodes chosen via \textsc{Select-Seed}.

\subsection{Effect of Out-degree on Network Diameter}
Extensive analyses \cite{hu2009evolution,mcglohon2011statistical,leskovec2005graphs} on evolving
real-world networks reveal two key temporal properties: network densification and diameter
shrinkage over time. Growth models can be adjusted to densify networks over time
by allowing node out-degree to increase super-linearly as a
function of network size. However, we lack a concrete understanding of existing
edge formation mechanisms' inability to preserve diameter shrinkage. Through our
analysis, we observe that the out-degree sequence of incoming nodes in
network models has a significant impact on effective diameter over time.
\begin{figure}
 \vspace{-10pt}
 \centering
 \includegraphics[width=\columnwidth]{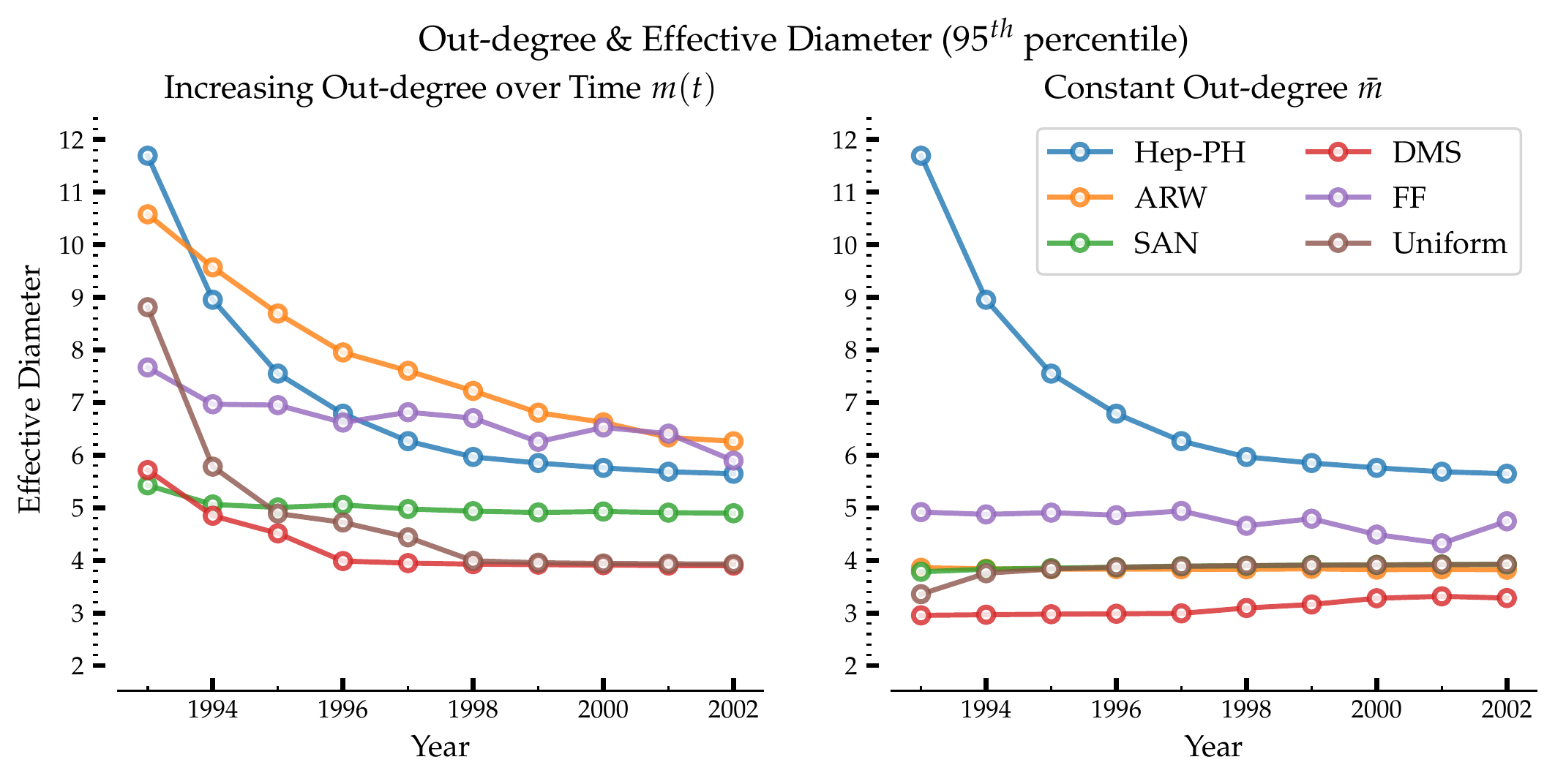}
 \caption{
 Effect of out-degree on effective diameter.
 The left subplot shows that increasing out-degree over time leads to diameter
 shrinkage for all models. The right subplot shows that
 constant out-degree sequence does not account for diameter
 shrinkage and consistently underestimates effective diameter of the \texttt{Hep-PH}
 network.}
 \label{fig:diameter}
 \vspace{-10pt}
\end{figure}

\Cref{fig:diameter} illustrates the effective diameter of network models fitted
to \texttt{Hep-PH} as a function of node out-degree sequence and time.
By increasing the
out-degree $m(t)$ over time using the method described
in~\cref{sub:Model Fitting}, network models representative of key edge formation
mechanisms---\texttt{ARW}, \texttt{FF}, \texttt{DMS} \&
\texttt{SAN}---generate networks that exhibit diameter shrinkage. In particular,
\texttt{FF} \footnote{\texttt{FF} inherently increases out-degree over time
because incoming nodes ``burn'' through the network for duration in proportion
to the network size.} and \texttt{ARW} mirror the observed rate at which the
effective diameter shrinks over time.
However, when the out-degree $\bar{m}=n^{-1}\sum_i m(i)$ of incoming nodes is constant,
fitted networks, including Forest Fire (\texttt{FF}), cannot preserve
shrinking diameter; the effective diameter of the fitted models remain
consistently lower than that of \texttt{Hep-PH}.

Increasing out-degree over time can effectively incorporate diameter shrinkage
in all representative network models. This phenomenon is best understood through
the simple \texttt{Uniform} null model, in which incoming nodes form edges to
existing nodes chosen uniformly at random. In the \texttt{Uniform} model, nodes with higher out-degree
have a greater probability of linking to existing
nodes that are structurally distant from each other. Consequently, over time,
incoming nodes with higher out-degree are more likely to bridge distant regions
of the network, reduce path length between existing nodes and subsequently
shrink effective diameter.

To summarize, our analysis indicates how increasing out-degree over time enables
existing models that rely on different edge formation processes to account for
diameter shrinkage.
% , an important temporal property of evolving real-world networks.

% \subsection{Connections to Vertex Copying}

\subsection{\texttt{ARW} Limitations}
We discuss three limitations of \texttt{ARW}. First, we  consider only
bibliographic  network datasets in which nodes form all edges at the time of joining.
This allows us to analyze edge formation in the absence of confounding edge
processes such as edge deletion and edge creation between existing nodes. We
plan to extend \texttt{ARW} to handle social networks, where individuals can
form edges at any time. One potential way is to incorporate random walks that
pause and resume intermittently, thus allowing for older nodes to connect with
more recent arrivals.
% Similarly, meta-path based random walks could potentially model interactions between nodes of different types in heterogeneous information networks.
Second, the out-degree $m(t)$ of incoming nodes in \texttt{ARW} rely on the
observed out-degree sequence, which might be unavailable in
datasets without fine-grained temporal data. In this case, \texttt{ARW}
can be adapted to rely on the prescribed range of densification exponent $\alpha_{\tiny \mbox{DPL}}$~\cite{leskovec2005graphs}
in real-world networks.
Since $e(t)=m(t)n(t)$, the power law relationship $e(t) \propto n(t)^{\alpha_{\tiny \mbox{DPL}}}$ between
number of edges $e(t)$ and nodes $n(t)$ at time $t$ implies that out-degree $m(t)$ must be proportional
to $n(t)^{\alpha_{\tiny \mbox{DPL}}-1}$.
Third, \texttt{ARW} focuses on modeling networks in which nodes have a single attribute.
The difficulty in incorporating multiple attributes into the edge formation mechanism rests
on how we measure attribute similarity. If two nodes are similar only when all their
attribute values are identical, we can simply create a new categorical attribute that encodes all
multiple attribute combinations and then directly apply \texttt{ARW}.
Additional analysis is necessary to identify definitions of attribute similarity that best
describe how multiple attributes influence individuals' edge formation processes.

\section{Related Work}
\label{sec:Related Work}
Network growth models seek to explain a subset of structural properties observed in real
networks.
% We point the reader to
% extensive surveys \cite{newman2003structure,albert2002statistical} of network growth models
% for more information.
Note that, unlike growth models, statistical models of network replication \cite{leskovec2010kronecker,pfeiffer2014attributed}
do not model how networks grow over time and are not relevant to our work.
Below, we discuss relevant and recent work on modeling network growth.

% PA
\textbf{Preferential Attachment \& Fitness}:
In preferential attachment and fitness-based models \cite{bell2017network,medo2011temporal,bianconi2001bose,caldarelli2002scale}, a new node $u$ links to an existing node $v$
with probability proportional to the attachment function $f(k_v)$, a function of
either degree $k_v$ or fitness $\phi_v$ of node $v$.
% Node fitness is defined as a dimensionless
% measure of node attractiveness.
For instance, linear preferential attachment functions
\cite{barabasi1999emergence,kumar2000stochastic,dorogovtsev2000structure} lead to
power law degree distributions and small diameter \cite{bollobas2004diameter}
and attachment functions of degree \& node age \cite{wang2013quantifying}
can preserve realistic temporal dynamics.
Extensions of preferential
attachment \cite{mossa2002truncation,zeng2005construction,wang2009local} that
incorporate resource constraints
disregard network properties other than power law degree distribution and small diameter.
Additional mechanisms are necessary to explain network properties
such as clustering and attribute mixing patterns.

\textbf{Triangle Closing}:
A set of models
\cite{holme2002growing,klemm2002highly,leskovec2008microscopic}
incorporate triadic closure using triangle closing mechanisms,
which increase {average} local clustering by forming edges between nodes
with one or more common neighbors. However, as explained in~\Cref{ss:tc}, models
based on preferential attachment and triangle closing do not preserve the local
clustering of low degree nodes.

\textbf{Attributed network models}:
These models \cite{de2013scale,karimi2017visibility,gong2012evolution,zheleva2009co}
account for the effect of attribute homophily on edge formation and preserve mixing patterns.
Existing models can be broadly categorized as (a) fitness-based model that define fitness as a function of
attribute similarity and (b) microscopic models of network evolution that require
complete temporal information about edge arrivals \& deletion. Our experiment
results in~\Cref{sub:Experimental Results} show that well-known attributed network models
\texttt{SAN} and \texttt{KA} preserve assortative
mixing patterns, degree distribution to some extent, but not local clustering
and degree-clustering correlation.

\textbf{Random walk models}:
% Random walk models, first introduced by Vazquez \cite{vazquez2000knowing,
% vazquez2003growing}, jointly explain the emergence of multiple properties
% observed in real networks under constraints of partial access and limited
% information.
First introduced by Vazquez \cite{vazquez2000knowing}, random walk models are inherently local.
Models \cite{blum2006random} in which
new nodes only link to terminal nodes of short random walks generate
networks with power law degree distributions \cite{chebolu2008pagerank} and
small diameter \cite{mehrabian2016sa} but do not preserve clustering. Models
such as \texttt{SK} \cite{saramaki2004scale}
and \texttt{HZ} \cite{herrera2011generating}, in which new nodes probabilistically link to
each visited nodes incorporate triadic closure but are not flexible enough to preserve
{skewed} local clustering of real-world networks, as shown in~\Cref{sub:Experimental Results}.
We also observe that recursive random walk models such as \texttt{FF} \cite{leskovec2005graphs}
preserve temporal properties such as shrinking diameter but considerably overestimate local clustering
and degree-clustering relationship of real-world networks.
Furthermore, existing random walk models disregard the effect of homophily and do not model attribute mixing
patterns.

% TODO: un-comment; commented out to fit paper to 10pages with author info
% \textbf{Recent Work}:
% P{\'a}lovics et al. \cite{palovics2017raising} use preferential and uniform
% attachment to model the decreasing power law exponent of real-world, undirected
% networks in which average degree increases over time. Singh et al.
% \cite{singh2017relay} (\texttt{RL}) augment preferential attachment to explain
% the shift in popularity of nodes over time via the concept of relay linking.
% Both models do not incorporate mechanisms to preserve clustering, attribute
% mixing patterns, and resource constraints that affect how individuals form edges
% in real-world networks.

To summarize, existing models do not explain how resource constrained and local processes
\textit{jointly} preserve multiple global network properties of attributed networks.

\vspace{-6pt}
\section{Conclusion}
\label{sec:Conclusion}
In this paper, we proposed a simple, interpretable model of attributed network
growth. \texttt{ARW} grows a directed network in the following manner: an
incoming node selects a seed node based on attribute similarity, initiates a
biased random walk to explore the network by navigating through neighborhoods of
existing nodes, and halts the random walk after connecting to a few visited
nodes. To the best of our knowledge, \texttt{ARW} is the first model that
unifies multiple sociological phenomena---bounded rationality; structural
constraints; triadic closure; attribute homophily; preferential
attachment---into a single local process to model global network structure
\textit{and} attribute mixing patterns.
We explored the parameter space of the model to show how each parameter
intuitively controls one or more key structural properties.
Our experiments on six
large-scale citation networks showed that \texttt{ARW} outperforms
relevant and recent existing models by a statistically significant
factor of 2.5--$10\times$.
We plan to extend the \texttt{ARW} model in two directions: modeling undirected, social
networks,  and analyzing the effect of attribute homophily
on the formation of temporal motifs \cite{paranjape2017motifs}

\vspace{-8pt}
% \clearpage
\bibliographystyle{ACM-Reference-Format}
\bibliography{draft}

\end{document}